\documentclass[11pt,preprint]{aastex}
\usepackage{natbib}
\tightenlines
\slugcomment{}

\newcommand \capsize {\footnotesize}

\newcommand \fion {f_{\rm ion}}
\newcommand \radpr {{\rm pr}}

\newcommand    \bB     {{\bf B}}
\newcommand    \br     {{\bf r}}
\newcommand    \by     {{\bf y}}

\newcommand	\beq	{\begin{equation}}
\newcommand	\beqa	{\begin{eqnarray}}

\newcommand	\cm	{\,{\rm cm}}

\newcommand	\eeq	{\end{equation}}
\newcommand	\eeqa	{\end{eqnarray}}
\newcommand	\erg	{\,{\rm ergs}}

\newcommand	\eV	{\,{\rm eV}}

\newcommand	\g	{\,{\rm g}}
\newcommand	\gtsim	{\gtrsim}		 

\newcommand	\He	{{\rm He}}

\newcommand     \IH     {I_{\rm H}}

\newcommand	\K	{\,{\rm K}}
\newcommand	\kms	{\,{\rm km~s}^{-1}}
\newcommand	\kpc	{\,{\rm kpc}}

\newcommand	\ltsim	{\lesssim}		 
\newcommand    \mH     {m_{\rm H}}

\newcommand    \Myr    {{\rm Myr}}

\newcommand    \nH     {n_{\rm H}}
\newcommand    \nrms   {n_{\rm rms}}
\newcommand	\pc	{\,{\rm pc}}

\newcommand	\s	{\,{\rm s}}

\newcommand    \taudrag {\tau_{\rm drag}}

\newcommand    \Volt   {{\rm V}}
\newcommand    \ymax   {y_{\rm max}}
\newcommand	\yr	{\,{\rm yr}}

\newcommand{\oldtext}[1]{}
\newcommand{\newtext}[1]{#1}

\countdef\decade=200
\advance\decade by \year
\countdef\hours=201
\advance\hours by \time
\divide\hours by 60
\countdef\mins=202
\advance\mins by \hours
\multiply\mins by 60
\multiply\hours by 100
\countdef\miltime=203
\advance\miltime by \hours
\advance\miltime by \time
\advance\miltime by -\mins

\begin{document}

\title{%
        \vspace*{-3.0em}
        {\normalsize\rm to appear in {\it The Astrophysical Journal}}\\ 
        \vspace*{1.0em}
 	On Radiation Pressure in Static, Dusty \ion{H}{2} Regions
	}

\author{B.T. Draine}
\affil{Princeton University Observatory, Peyton Hall, Princeton,
       NJ 08544; {\tt draine@astro.princeton.edu}}

\begin{abstract}
Radiation pressure acting on gas and dust causes \ion{H}{2} regions to have
central densities that are lower than the density near the ionized
boundary.  \ion{H}{2} regions in static equilibrium comprise a 
family of similarity solutions with 3 parameters:
$\beta$, $\gamma$, and the product $Q_0\nrms$; $\beta$ characterizes the
stellar spectrum, $\gamma$ characterizes the dust/gas ratio,
$Q_0$ is 
the stellar ionizing output (photons/s), and
$\nrms$ is the rms density within the ionized region.
Adopting standard values for $\beta$ and $\gamma$, varying $Q_0\nrms$
generates a one-parameter family of density profiles, ranging from
nearly uniform-density (small $Q_0\nrms$) to
shell-like (large $Q_0\nrms$).
When $Q_0\nrms\gtsim 10^{52}\cm^{-3}\s^{-1}$,
dusty \ion{H}{2} regions have conspicuous central cavities, even if no
stellar wind is present.
For given $\beta$, $\gamma$ and $Q_0\nrms$, a fourth quantity, which can
be $Q_0$, determines the overall size and density of the
\ion{H}{2} region.
Examples of density and emissivity profiles are given.
We show how quantities of interest 
--
such as the peak-to-central emission measure ratio, 
the rms-to-mean density ratio,
the edge-to-rms density ratio, and the fraction of the
ionizing photons absorbed by the gas 
--
depend on 
$\beta$, $\gamma$, and $Q_0\nrms$.
For dusty \ion{H}{2} regions, compression of the gas and 
dust into an ionized shell
results in a substantial increase in the fraction of the stellar photons
that actually ionize H
(relative to a uniform density \ion{H}{2} region with the same dust/gas
ratio and density $n=\nrms$).
We discuss the extent to which
radial drift of dust grains in \ion{H}{2} regions can alter the
dust-to-gas ratio.
The applicability of these solutions 
to real \ion{H}{2} regions is discussed.
\end{abstract}


\section{Introduction
         \label{sec:intro}}
\citet{Stromgren_1939} idealized photoionized nebulae around
hot stars as static, spherical regions with
a uniform density of ionized gas out to a bounding radius $R$.
The Str\"omgren sphere model continues 
to serve as the starting point for studies of \ion{H}{2} regions
around hot stars.  
However, a number of physical effects
lead to departures from the simple Str\"omgren sphere model:
dynamical expansion of the \ion{H}{2} region if the pressure in the surrounding
neutral medium cannot confine the ionized gas; 
deviations from sphericity due to nonuniform density;
motion of the star relative to the gas; 
injection of energy and momentum by a stellar wind; 
absorption of H-ionizing photons by dust grains; 
and radiation pressure acting on gas and dust.  
Each of these effects has been the object
of a number of investigations, beginning with the study of
ionization fronts by \citet{Kahn_1954}.  

\citet{Savedoff+Greene_1955} appear to have been the first to discuss
the expansion of a spherical \ion{H}{2} region in an initially uniform
neutral medium.
\citet{Mathews_1967,Mathews_1969} and 
\citet{Gail+Sedlmayr_1979}
calculated the dynamical expansion of
an \ion{H}{2} region produced by an O star in a medium that was initially
neutral, including the effects of radiation
pressure acting on the dust.
\citet{Mathews_1967,Mathews_1969}
showed that radiation pressure on dust
would produce low-density central cavities in \ion{H}{2} regions.  
More
recently, \citet{Krumholz+Matzner_2009} reexamined the role of
radiation pressure on the expansion dynamics of \ion{H}{2} regions,
concluding that radiation pressure 
is generally unimportant for \ion{H}{2} regions ionized
by a small number of stars, but is important for 
\newtext{the expansion dynamics of}
giant \ion{H}{2} regions
surrounding clusters containing many O-type stars.
Their study concentrated on the forces acting on the dense shell
of neutral gas and dust bounding the \ion{H}{2} region, hence they
did not consider the density structure within the ionized region.

Dust absorbs
$h\nu > 13.6\eV$ photons that would otherwise be able to ionize hydrogen,
thereby reducing the extent of the ionized zone.
\citet{Petrosian+Silk+Field_1972} developed analytic approximations for
dusty \ion{H}{2} regions.
They assumed the gas density to be uniform, with a constant dust-to-gas
ratio, and found that dust could absorb a substantial fraction of the
ionizing photons in dense \ion{H}{2} regions.
Petrosian et al.\ 
did not consider the effects of radiation pressure.

\citet{Dopita+Groves+Sutherland+Kewley_2003,
       Dopita+Fischera+Crowley+etal_2006}
constructed models of compact \ion{H}{2} regions, 
including the effects of radiation pressure
on dust, and presented models for different ionizing stars
and bounding pressures.  In these models, radiation pressure produces
a density gradient within the ionized gas.

The present paper provides a systematic discussion of 
the structure of dusty \ion{H}{2} regions that are
assumed to be in equilibrium with an external bounding pressure.
The assumptions and governing equations are presented in 
\S\ref{sec:equilibrium}, where it is shown that dusty \ion{H}{2}
regions are essentially described by a 3-parameter family 
of similarity solutions.
In \S\ref{sec:results} we show density profiles for selected cases,
as well as surface brightness
profiles.  The characteristic ionization parameter
$U_{1/2}$ and the fraction $(1-\fion)$
of the ionizing photons absorbed by dust are calculated.
Dust grain drift is examined in \S\ref{sec:dust drift}, where it is shown that
it can alter the dust-to-gas ratio in the centers of high
density \ion{H}{2} regions.
The results are discussed in \S\ref{sec:discussion}, and summarized in
\S\ref{sec:summary}.
	
\section{\label{sec:equilibrium}
          Equilibrium Model}

Consider the idealized problem of a static, spherically-symmetric
equilibrium \ion{H}{2} region
ionized by a point source, representing either a single star or
a compact stellar cluster.
Assume a constant
dust-to-gas ratio (the validity of this assumption will be examined
later).
For simplicity, ignore scattering, and assume $\sigma_d$, the dust
absorption cross section per H nucleon, to be independent of
photon energy $h\nu$
over the $\sim 5\eV$ to $\sim30\eV$ range containing most of the
stellar power.

Let the star have luminosity $L=L_n+L_i=L_{39}10^{39}\erg\s^{-1}$,
where $L_n$ and $L_i$
are the luminosities in $h\nu < 13.6\eV$ and $h\nu > 13.6\eV$ photons,
respectively.
The rate of emission of $h\nu>13.6\eV$ photons is 
$Q_0\equiv 10^{49}Q_{0,49}\s^{-1}$ 
and the mean energy of the ionizing photons is
$\langle h\nu\rangle_i\equiv L_i/Q_0$.
A single main sequence star of spectral type O6V
has $L_{39}=0.80$ and $Q_{0,49}=0.98$
\citep{Martins+Schaerer+Hillier_2005}.
A compact cluster of OB stars might be treated as a point source with
much larger values of $Q_{0,49}$ and $L_{39}$.

Ignore He, and assume the H to be nearly fully ionized, with
photoionization balancing ``Case B'' radiative recombination,
with ``on-the-spot'' absorption of $h\nu>13.6\eV$ recombination radiation.
Take the effective radiative recombination coefficient to be
$\alpha_B\approx2.56\times10^{-13}T_4^{-0.83}\cm^3\s^{-1}$ for
$0.5\ltsim T_4\ltsim 2$, with $T_4\equiv T/10^4\K$, where $T$ is
the gas temperature.

Assume the gas to be in dynamical equilibrium (the neutral gas outside
the ionized zone is assumed to provide a confining pressure).
Static equilibrium then requires that the force per unit volume
from radiation pressure be balanced by the pressure gradient:
\beq \label{eq:dynamical equilibrium}
n\sigma_d \frac{\left[L_n e^{-\tau}+L_i \phi(r)\right]}{4\pi r^2 c} + 
\alpha_B n^2\frac{\langle h\nu\rangle_i}{c} - 
\frac{d}{dr}\left(2nkT\right)
= 0 ~~~,
\eeq
where $n(r)$ is the proton density,
$L_i \phi(r)$ is the power in $h\nu>13.6\eV$ photons crossing a sphere
of radius $r$, and $\tau(r)$ is the dust absorption optical depth.
Eq.\ (\ref{eq:dynamical equilibrium}) underestimates 
the radiation pressure force, because it assumes that recombination radiation
(including Lyman-$\alpha$)
and cooling radiation escape freely.

The functions $\phi(r)$ and $\tau(r)$ are determined by
\beqa \label{eq:dphi/dr}
\frac{d\phi}{dr} &=& 
-\frac{1}{Q_0} \alpha_B n^2 4\pi r^2
- n\sigma_d \phi ~~~,~~~
\\ \label{eq:dtau/dr}
\frac{d\tau}{dr} &=& n\sigma_d ~~~,
\eeqa
with boundary conditions $\phi(0)=1$ and $\tau(0)=0$.
Define a characteristic density and length scale
\beqa \label{eq:define n_0}
n_0 
&\equiv& \frac{4\pi\alpha_B}{Q_0}
\left(\frac{2ckT}{\alpha_B\langle h\nu\rangle_i}\right)^3 
= 
4.54\times10^5 ~\frac{T_4^{4.66}}{Q_{0,49}}
\left(\frac{18\eV}{\langle h\nu\rangle_i}\right)^3\cm^{-3}
~~~,~~~
\\
\lambda_0 
&\equiv& \frac{Q_0}{4\pi\alpha_B}
\left(\frac{\alpha_B\langle h\nu\rangle_i}{2ckT}\right)^2
=
2.47\times10^{16}~\frac{Q_{0,49}}{T_4^{2.83}}
\left(\frac{\langle h\nu\rangle_i}{18\eV}\right)^2
\cm
~~~,~~~
\eeqa
and the dimensionless parameters
\beqa \label{eq:define beta}
\beta &\equiv& \frac{L_n}{L_i} = \frac{L}{L_i} - 1 
= 3.47~\frac{L_{39}}{Q_{0,49}}
\left(\frac{18\eV}{\langle h\nu\rangle_i}\right) -1
~~~,~~~
\\ \label{eq:define gamma}
\gamma
&\equiv& \left(\frac{2ckT}{\alpha_B \langle h\nu\rangle_i}\right)
\sigma_d
= 11.2 ~T_4^{1.83}\left(\frac{18\eV}{\langle h\nu\rangle_i}\right)
\left(\frac{\sigma_d}{10^{-21}\cm^2}\right)
~~~.~~~
\eeqa

The parameter $\beta$, the
ratio of the power in non-ionizing photons to the power in photons
with $h\nu>13.6\eV$, depends solely on the stellar spectrum.
We take $\beta=3$ as our standard value, corresponding to the
spectrum of a $T_\star=32000\K$ blackbody, but we also
consider $\beta=2$ ($T_\star=45000\K$) and $\beta=5$ ($T_\star=28000\K$);
the latter value might apply to a cluster of O and B stars.

\begin{figure}[hbt]
\begin{center}
\includegraphics[width=10cm,angle=270]
   {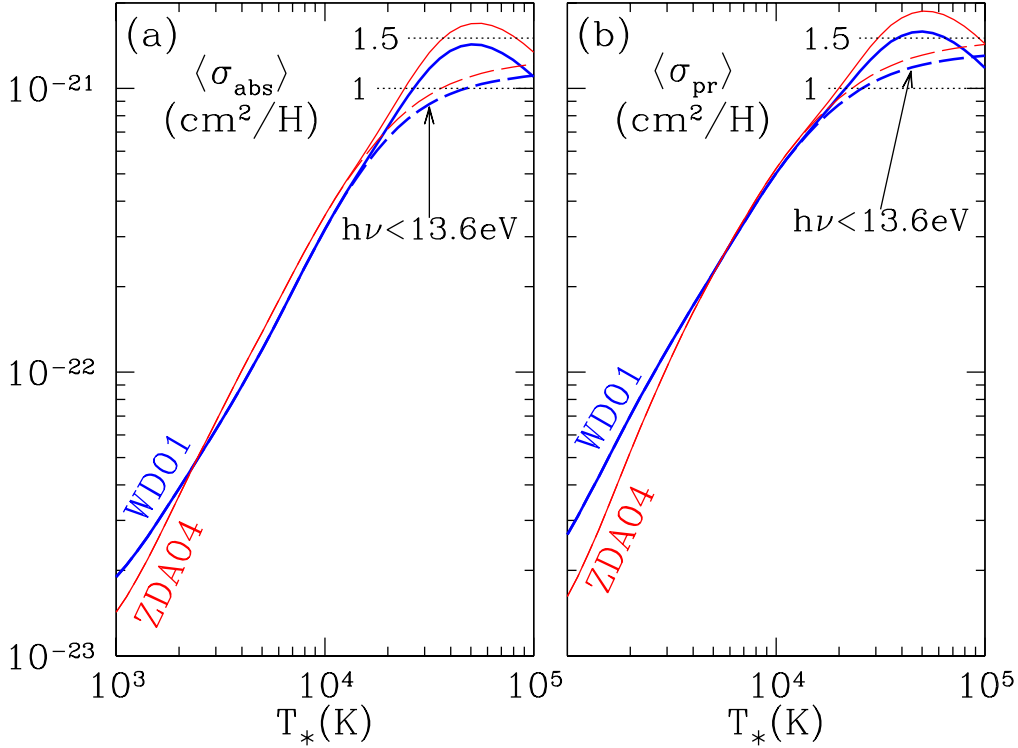}
\caption{\label{fig:sigma_rp}
   \capsize
   (a) Absorption cross section per H, and 
   (b) radiation pressure cross section per H, averaged over 
   blackbody spectra, as functions
   of the blackbody temperature $T_\star$, for the
   dust models of \citet[][WD01]{Weingartner+Draine_2001a} and
   \citet[][ZDA04]{Zubko+Dwek+Arendt_2004}.
   Broken lines show averages over $h\nu<13.6\eV$ only, appropriate
   for dust in neutral gas.
   }
\end{center}
\end{figure}


Momentum can be transferred to a dust grain by photon absorption, but also
by scattering.
The cross section $\sigma_d$
appearing in eq.\ (\ref{eq:dynamical equilibrium})
should be $\langle\sigma_\radpr\rangle$, the
radiation pressure cross section per H,
$\sigma_\radpr(\nu)
\equiv
\sigma_{\rm abs} + (1-\langle\cos\theta\rangle)\sigma_{\rm sca}$,
averaged over
the spectrum of the radiation field at radius $r$, where
$\sigma_{\rm abs}(\nu)$ and $\sigma_{\rm sca}(\nu)$ are the absorption
and scattering cross section per H, and $\langle\cos\theta\rangle$ is
the mean value of the cosine of the scattering angle $\theta$ for photons
of frequency $\nu$.

In eqs.\ (\ref{eq:dphi/dr}) and (\ref{eq:dtau/dr}),
$\sigma_d$
characterizes the effectiveness of the dust in attenuating
the radiation field.
While scattering does not destroy the photon, it does increase the
probability of the photon undergoing subsequent absorption.
Thus, the value of $\sigma_d$ in eqs.\ (\ref{eq:dphi/dr}) and (\ref{eq:dtau/dr})
should exceed $\langle\sigma_{\rm abs}\rangle$.

Figure \ref{fig:sigma_rp}(a) shows the dust absorption cross section per H
nucleon averaged over a blackbody spectrum, for two dust models 
\citep{Weingartner+Draine_2001a, Zubko+Dwek+Arendt_2004} that
reproduce the wavelength-dependent extinction in the diffuse interstellar
medium using mixtures of PAHs, graphite, and amorphous silicate grains.
Fig.\ \ref{fig:sigma_rp}(b)
shows that $\langle\sigma_\radpr\rangle$,
the radiation pressure cross section 
averaged over blackbody spectra,
is only slightly larger than $\langle\sigma_{\rm abs}\rangle$.
Given the uncertainties in the nature of the dust in \ion{H}{2}
regions, it is reasonable to ignore the distinction between
$\langle\sigma_\radpr\rangle$ and the attenuation cross section
and simply take $\sigma_d=\langle\sigma_\radpr\rangle$ in
eq.\ (\ref{eq:dynamical equilibrium}--\ref{eq:dtau/dr}).

For dust characteristic of the diffuse ISM, one could
take 
$\langle\sigma_\radpr\rangle\approx1.5\times10^{-21}\cm^2{\,\rm H}^{-1}$ for 
$2.5\times10^4\K\ltsim T_{\rm rad}\ltsim 5\times10^4\K$.
However, dust within an \ion{H}{2} region 
may differ from average interstellar dust.
For example, the small-size end of the size distribution might be
suppressed, in which case $\sigma_d$ would be reduced.
Low metallicity galaxies will also have lower values of $\sigma_d$, simply
because there is less material out of which to form grains.
In the present work we will assume a factor $\sim$1.5 reduction in $\sigma_d$
relative to the local diffuse ISM, taking 
$\sigma_d\approx 1\times10^{-21}\cm^2{\,\rm H}^{-1}$ 
as the nominal value, but larger and smaller values of
$\sigma_d$ will also be considered.

The dimensionless parameter $\gamma$ (defined in eq.\ \ref{eq:define gamma})
depends also on the gas temperature $T$ and on the
mean ionizing photon energy $\langle h\nu\rangle_i$, but these are not likely
to vary much for \ion{H}{2} regions around OB stars.
We take $\gamma=10$ as a standard value,
but will also consider $\gamma=5$ and
$\gamma=20$.
Low-metallicity systems would be
characterized by small values of $\gamma$.

Switching to dimensionless variables 
$y\equiv r/\lambda_0$, $u\equiv n_0/n$, the governing equations
(\ref{eq:dynamical equilibrium}--\ref{eq:dtau/dr})
become
\beqa \label{eq:dudy}
\frac{du}{dy} &=& -1 - \gamma\left(\beta e^{-\tau} + \phi\right)\frac{u}{y^2}
~~~,~~~
\\ \label{eq:dphidy}
\frac{d\phi}{dy} &=& -\frac{y^2}{u^2} - \gamma\frac{\phi}{u}
~~~,~~~
\\ \label{eq:dtaudy}
\frac{d\tau}{dy} &=& \frac{\gamma}{u} ~~~,
\eeqa
with initial conditions 
$\phi(0) = 1$ and
$\tau(0) = 0$.
The solutions are defined for $0<y\leq\ymax$, where $\ymax$ is
determined by the boundary condition $\phi(\ymax)=0$.
The actual radius of the ionized zone is $R=\ymax\lambda_0$.
For each solution $u(y)$
the mean density is
\beq
\langle n \rangle = \frac{3n_0}{\ymax^3}\int_0^{\ymax} \frac{1}{u} y^2 dy
~~~,
\eeq%
the root-mean-square density is
\beq
\nrms \equiv 
n_0 
\left[\frac{3}{\ymax^3}
\int_0^{\ymax} \frac{1}{u^2}y^2 dy\right]^{1/2}
~~~,
\eeq
and the gas pressure at the edge of the \ion{H}{2} region is
\beq
p_{\rm edge} = 2 n(R) kT = \frac{2n_0 kT}{u(\ymax)}
~~~.
\eeq%
Let 
\beq
R_{s0}\equiv \left(\frac{3Q_0}{4\pi\nrms^2\alpha_B}\right)^{1/3}
= 2.10\times10^{18} \frac{Q_{0,49}^{1/3}}{n_{\rm rms,3}^{2/3}}
T_4^{0.28}\cm
~~~
\eeq
be the radius of a dustless Str\"omgren sphere with density 
$\nrms = 10^3 n_{{\rm rms},3}\cm^{-3}$.
The fraction of the $h\nu>13.6\eV$
photons that are absorbed by H is simply
\beq
\fion = \left(\frac{R}{R_{s0}}\right)^3 ~~~.~~~
\eeq 

For given $(\beta,\gamma)$, varying the initial value\footnote{
   For $\gamma>0$, $u\propto \exp[(\beta+1)\gamma/y]\rightarrow \infty$ 
   as $y\rightarrow 0$, and the integration must start at
   some small $y>0$.} 
of $u=n_0/n$ at some fixed $y=r/\lambda_0$
generates solutions with different density profiles.
Therefore the full set of solutions forms a three-parameter family
of similarity solutions $u(y)$, $\phi(y)$, and $\tau(y)$, 
parametrized
by $\beta$, $\gamma$, and a third parameter.
The third parameter can be taken to be $Q_0\nrms$.
For dusty \ion{H}{2} regions, an alternative choice for the
third parameter is the dust optical depth on a path $R_{s0}$
with density $\nrms$:
\beqa \label{eq:taud0}
\tau_{d,0} \equiv \nrms R_{s0} \sigma_d &=& 
2.10\left(Q_{0,49}n_{\rm rms,3}\right)^{1/3} 
T_4^{0.28}
\frac{\sigma_d}{10^{-21}\cm^2}
\\ 
&=& 0.188 \gamma \left(Q_{0,49}n_{\rm rms,3}\right)^{1/3}
      T_4^{-1.55}
      \frac{\langle h\nu\rangle_i}{18\eV}
~~~.
\eeqa

The static \ion{H}{2} regions described by 
eq.\ (\ref{eq:dynamical equilibrium}\,--\,\ref{eq:dtau/dr})
are determined by 7 distinct dimensional quantities:
three parameters describing the central star 
($Q_0$, $\langle h\nu\rangle_i$, and $L_n$),
the recombination rate coefficient $\alpha_B$,
the thermal energy $kT$,
the dust cross section per nucleon $\sigma_d$,
and the external pressure $p_{\rm edge}$ confining the \ion{H}{2}
region.
According to the present analysis, this 7-parameter family of solutions
actually reduces to a 3-parameter family of similarity solutions.
The dimensionless parameters $\beta$ and $\gamma$, plus choice of an
initial value for
the function $u$ near $y=0$, suffice to determine the scaled density
profile
$n(r)/n_0$ and radius $\ymax=R/\lambda_0$: this is the 3-parameter
family of similarity solutions.

Specifying numerical values for the ratios $Q_0/\alpha_B$ and
$kT/(\alpha_B\langle h\nu\rangle_i)$ fixes the values of $n_0$ and
$\lambda_0$, thus giving $n(r)$ for $r<R$.
Thus far we have invoked 5 independent parameters, but have not actually
specified either $kT$ or $\alpha_B$.

Specifying $kT$
and $\alpha_B$ -- the 6th and 7th parameters -- allows us to
compute the actual values of $Q_0$ and $\langle h\nu\rangle_i$,
and the bounding pressure $p_{\rm edge}=2n(R)kT$.

If the ``initial value'' of $u$ near the origin is taken as a
boundary condition, then
$p_{\rm edge}$ emerges as a derived quantity.  However, if we
instead regard $p_{\rm edge}$ as a boundary condition, 
then the initial value of
$u$ ceases to be a free parameter, and instead must be found 
(e.g., using  a shooting technique) so as to give the
correct boundary pressure $p_{\rm edge}$: the initial value of $u$
is thus determined by the
7 physical parameters ($Q_0$, $\langle h\nu\rangle_i$, $L_n$,
$\alpha_B$, $T$, $\sigma_d$, and $p_{\rm edge}$).

Thus we see how the 3 parameter family of dimensionless similarity
solutions corresponds to a 7 parameter family of physical solutions.

\begin{figure}[hbt]
\begin{center}
\includegraphics[width=6cm,angle=270]%
   {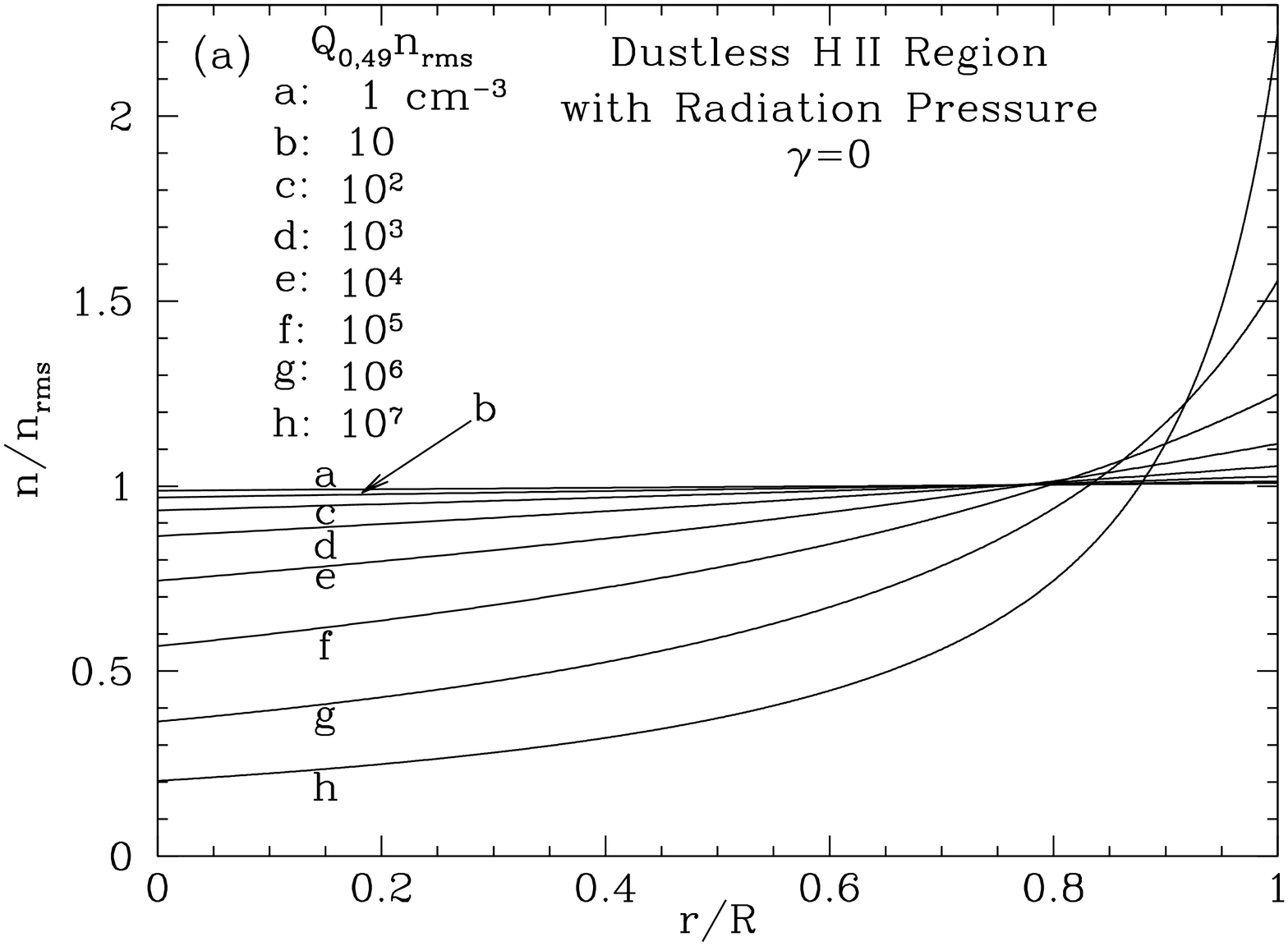}
\includegraphics[width=6cm,angle=270]%
   {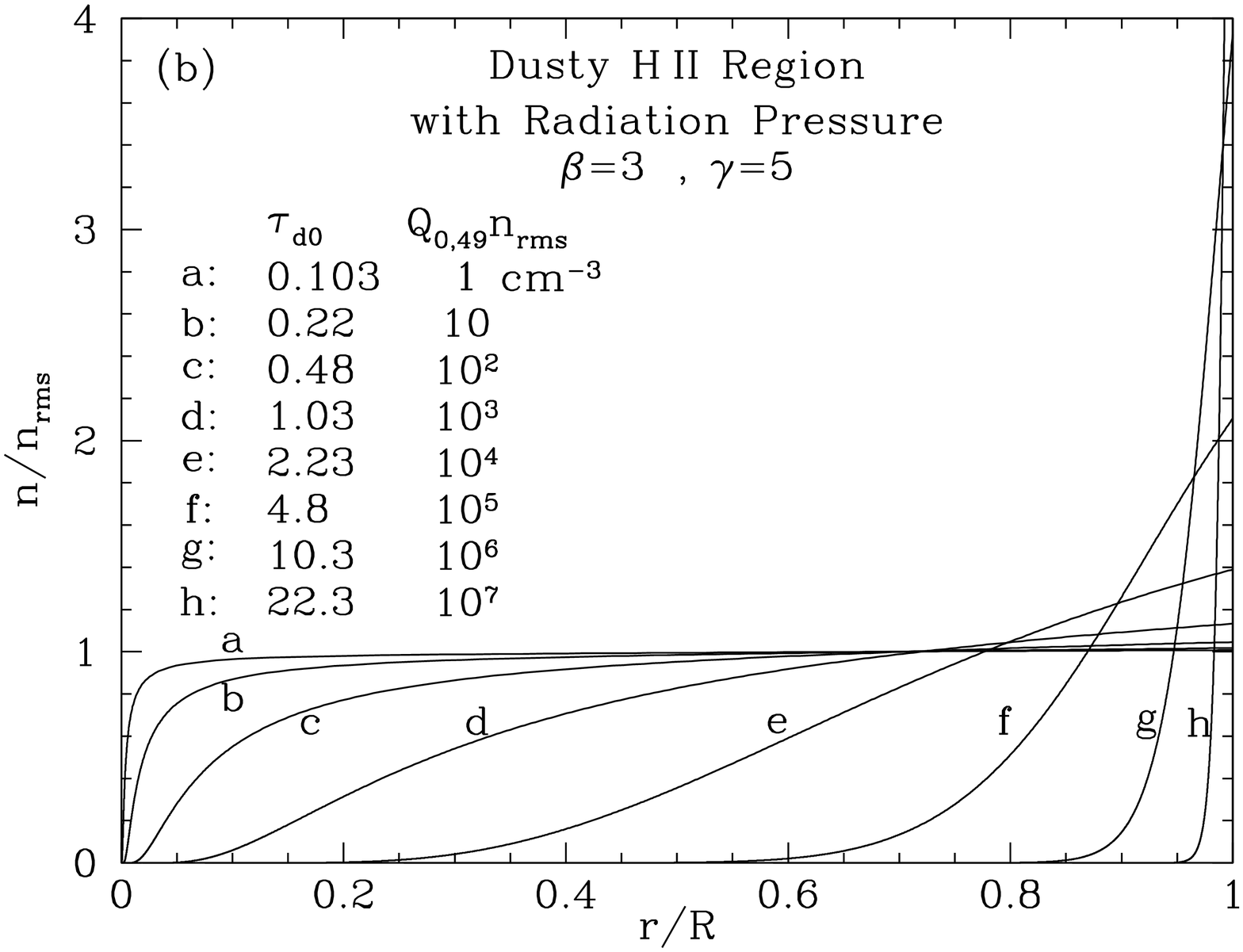}
\includegraphics[width=6cm,angle=270]%
   {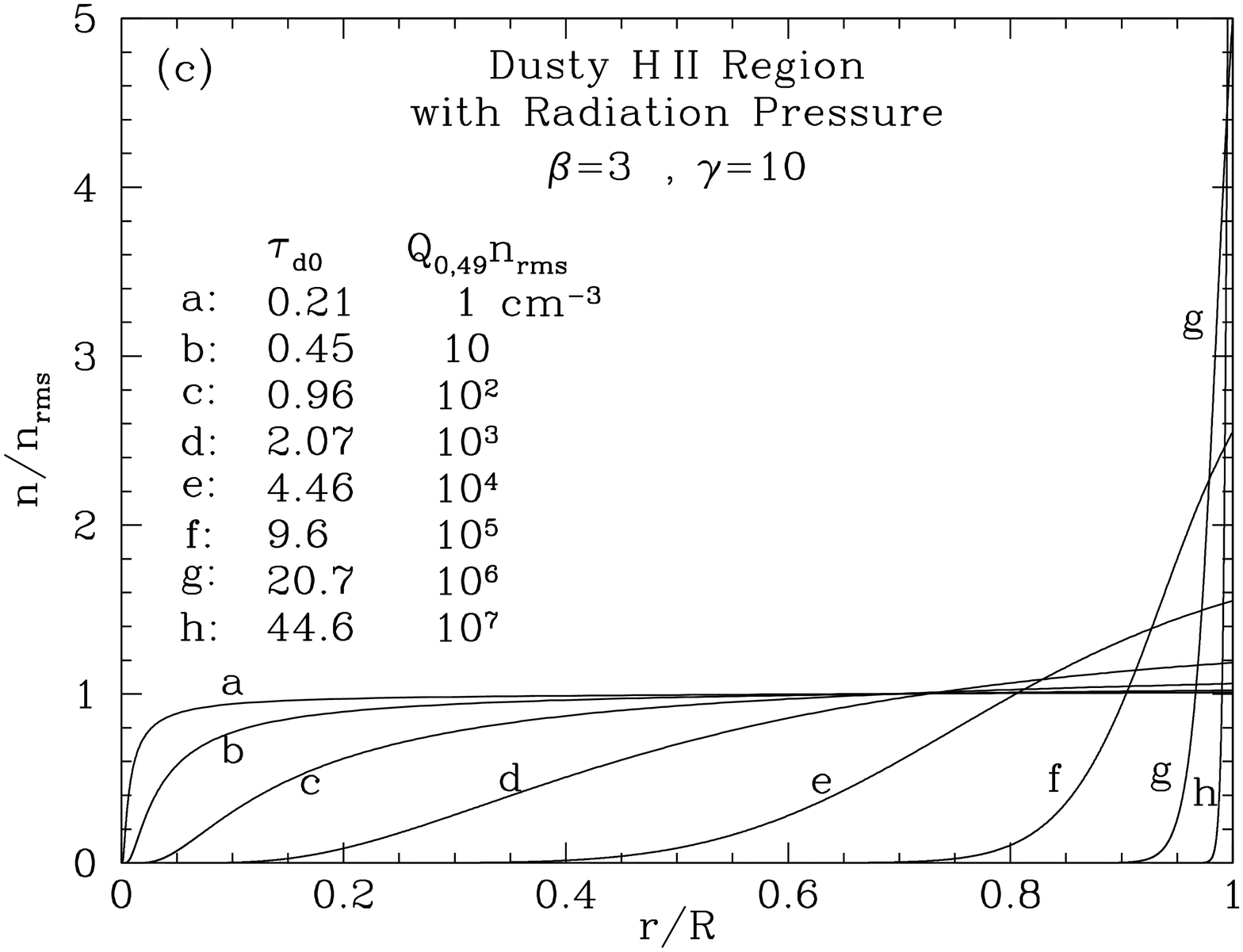}
\includegraphics[width=6cm,angle=270]%
   {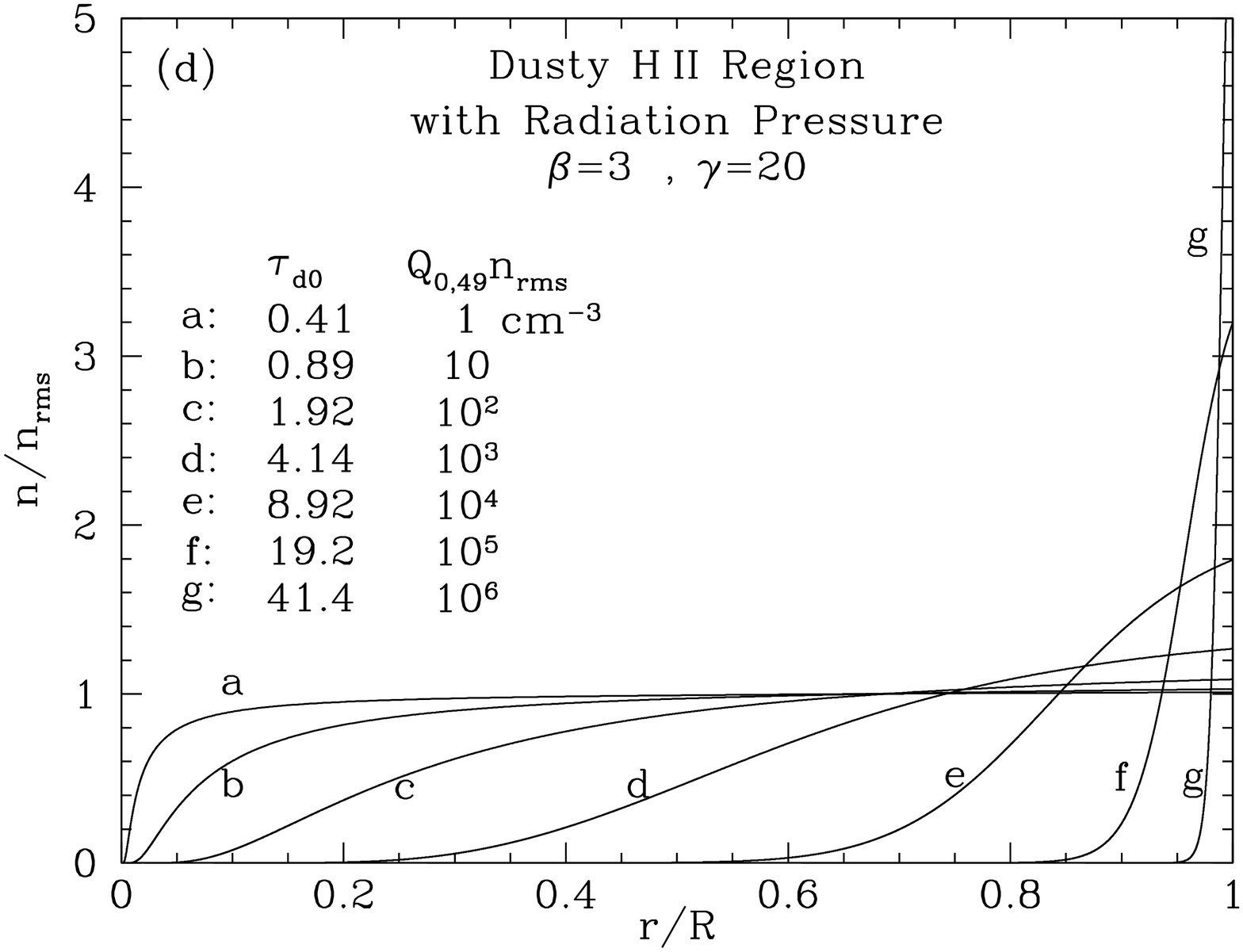}
\caption{\label{fig:nprofs gamma=0, 5, 10, 20}
   \capsize
   Normalized density profiles of static equilibrium \ion{H}{2} regions, as
   a function of $r/R$, where $R$ is the radius of the ionized
   region.  Profiles are shown for 7 values of
   $Q_0\nrms$; the numerical values given in the legends assume 
   $T_4=0.94$ and $\langle h\nu\rangle_i=18\eV$.
   (a) Dustless ($\gamma=0$);
   (b) $\gamma=5$;
   (c) $\gamma=10$;
   (d) $\gamma=20$.
   }
\end{center}
\end{figure}
\begin{figure}[hbt]
\begin{center}
\includegraphics[width=10cm,angle=0]%
   {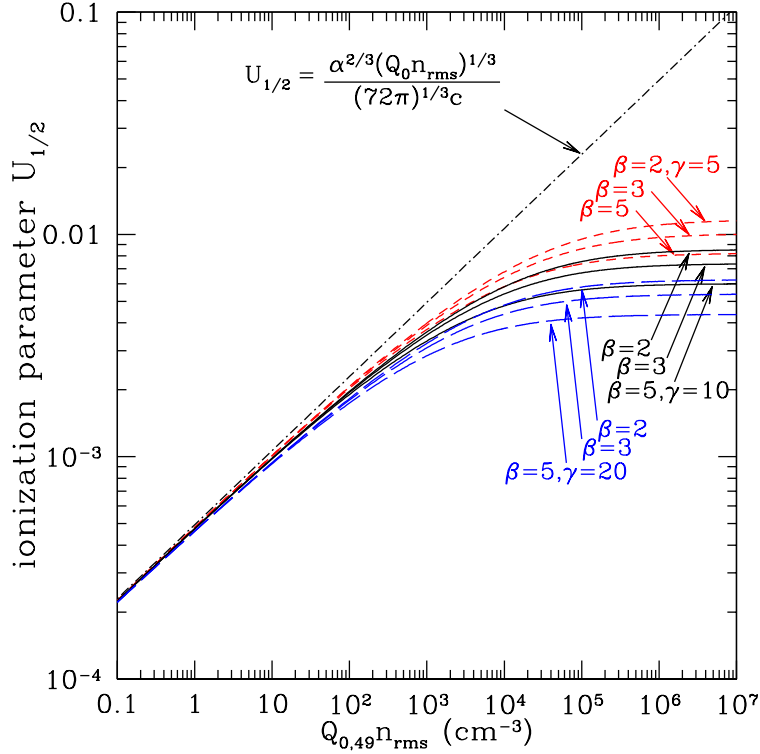}
\caption{\label{fig:U}
   \capsize
   Ionization parameter $U_{1/2}$ at the half-ionization
   radius in dusty H\,II regions (see text),
   calculated assuming $T_4=0.94$ and $\langle h\nu\rangle_i=18\eV$.
   }
\end{center}
\end{figure}

\section{\label{sec:results}
         Results}

Equations (\ref{eq:dudy}-\ref{eq:dtaudy}) can be integrated numerically.
Figure \ref{fig:nprofs gamma=0, 5, 10, 20}a shows the solution for the
case where no dust is present ($\gamma=0$).
Radiation pressure associated with photoionization produces a
density gradient in the \ion{H}{2} region, but it is
modest unless $Q_0\nrms$ is very large.
The central density is nonzero.  
For 
$Q_{0,49}\nrms\ltsim 10^3\cm^{-3}$, 
the density is uniform to within $\pm15\%$.

As discussed above, 
the dust abundance relative to H is characterized by the parameter
$\gamma$.
Density profiles are shown in 
Fig.\ \ref{fig:nprofs gamma=0, 5, 10, 20}b-d for $\beta=3$ and
$\gamma=5, 10, 20$, corresponding approximately to 
$\sigma_d=0.5, 1, 2 \times10^{-21}\cm^2$.
With dust present, 
the density formally goes to zero at $r=0$.
For fixed $\gamma$,
the size of the low-density central cavity 
(as a fraction of the radius $R$ of the ionization front) 
increases with increasing $Q_0\nrms$.
The enhancement of the density near the ionization front also becomes more
pronounced as $Q_0\nrms$ is increased.
For 
$\beta=3$, $\gamma=10$ and $Q_{0,49}\nrms=10^5\cm^{-3}$, 
we find $n(R)=2.5\nrms$.

\begin{figure}[bt]
\begin{center}
\includegraphics[width=6cm,angle=270]%
   {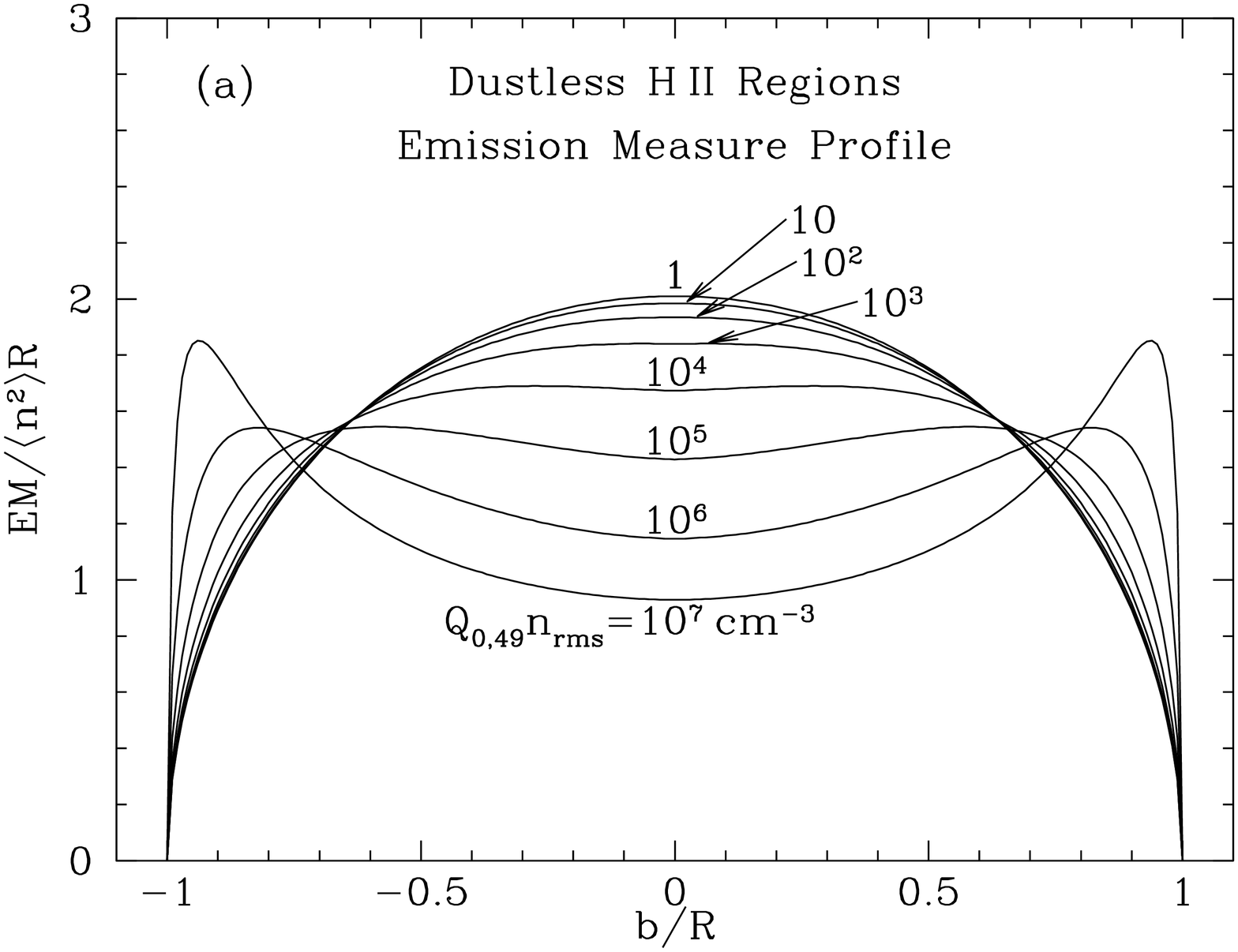}
\includegraphics[width=6cm,angle=270]%
   {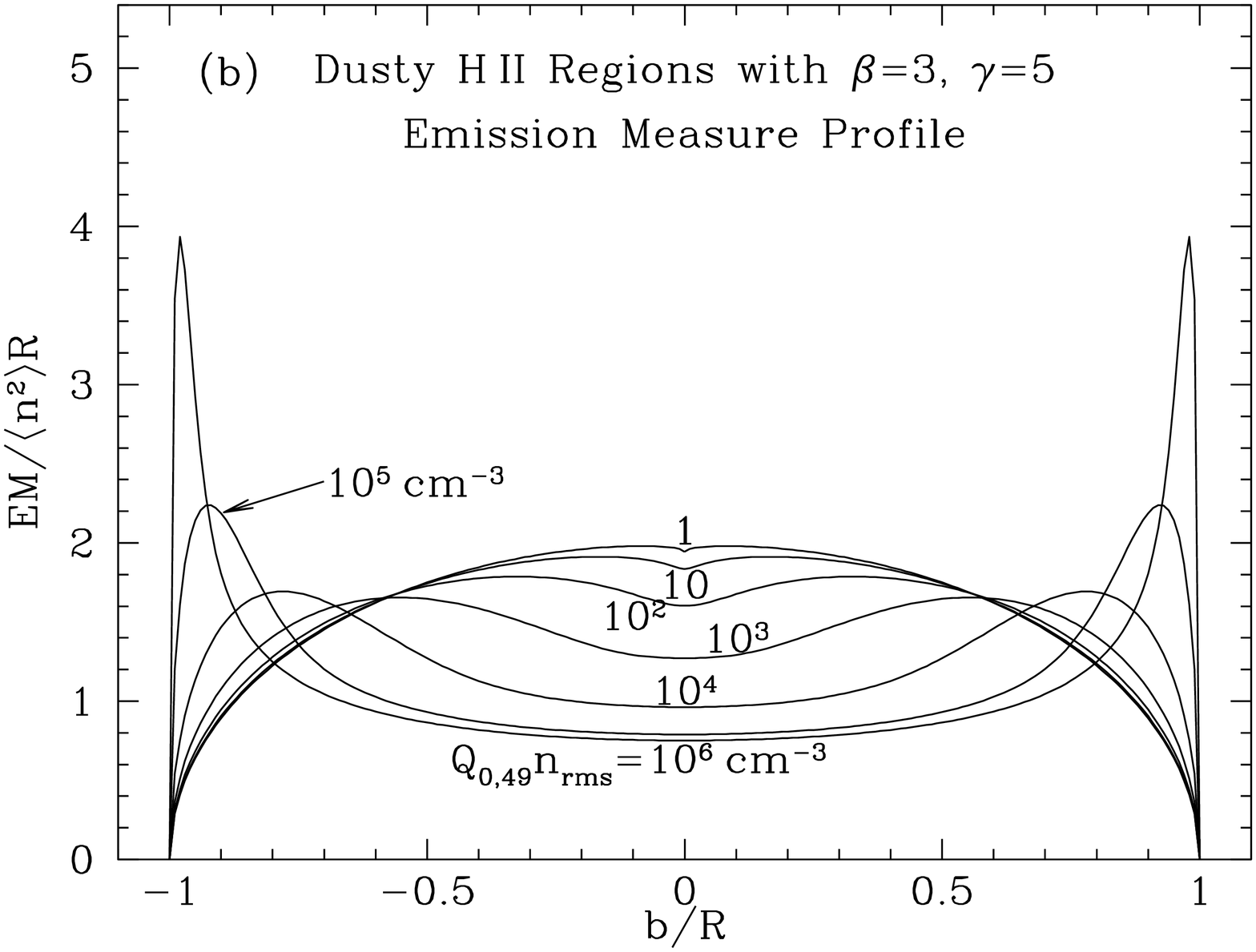}
\includegraphics[width=6cm,angle=270]%
   {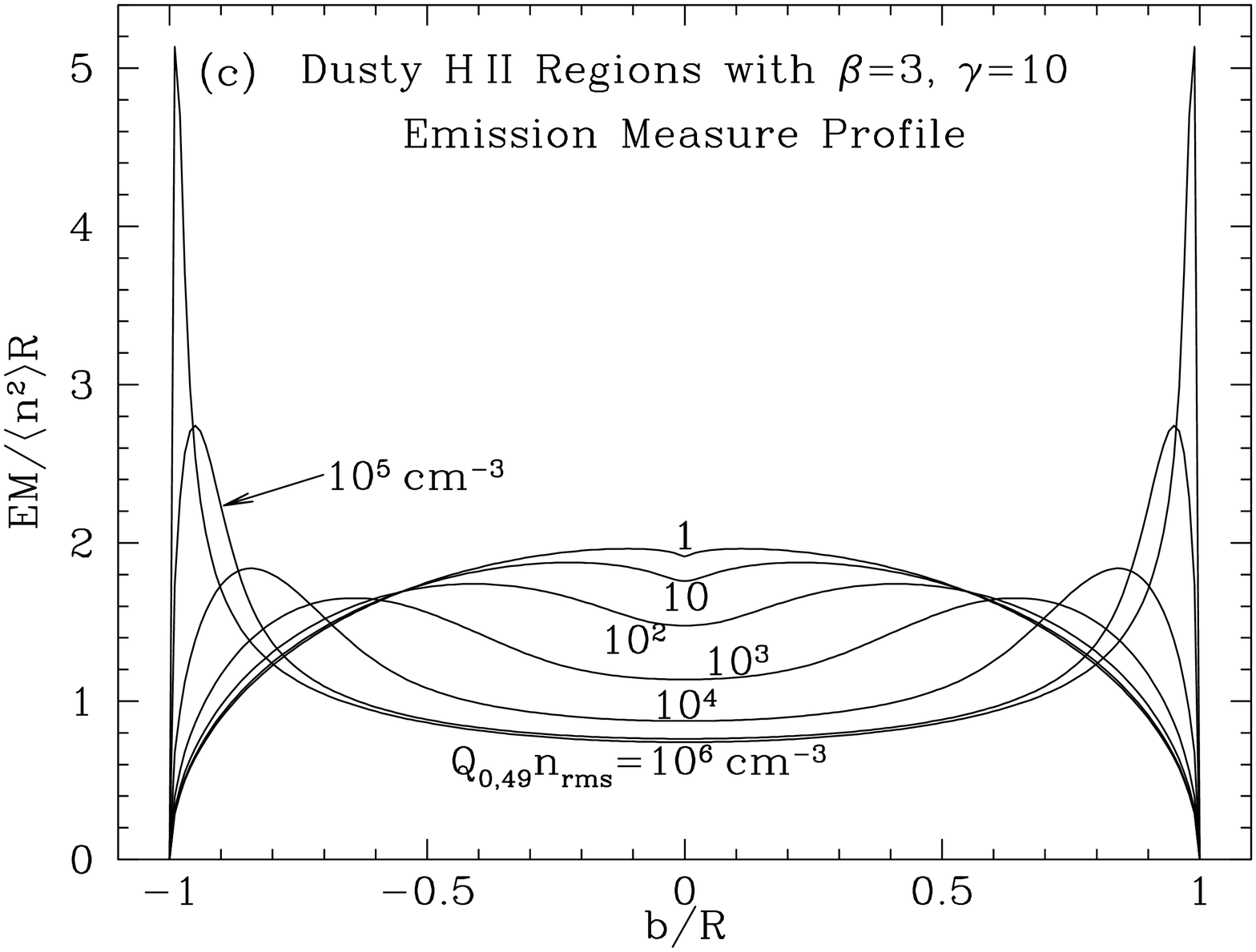}
\includegraphics[width=6cm,angle=270]%
   {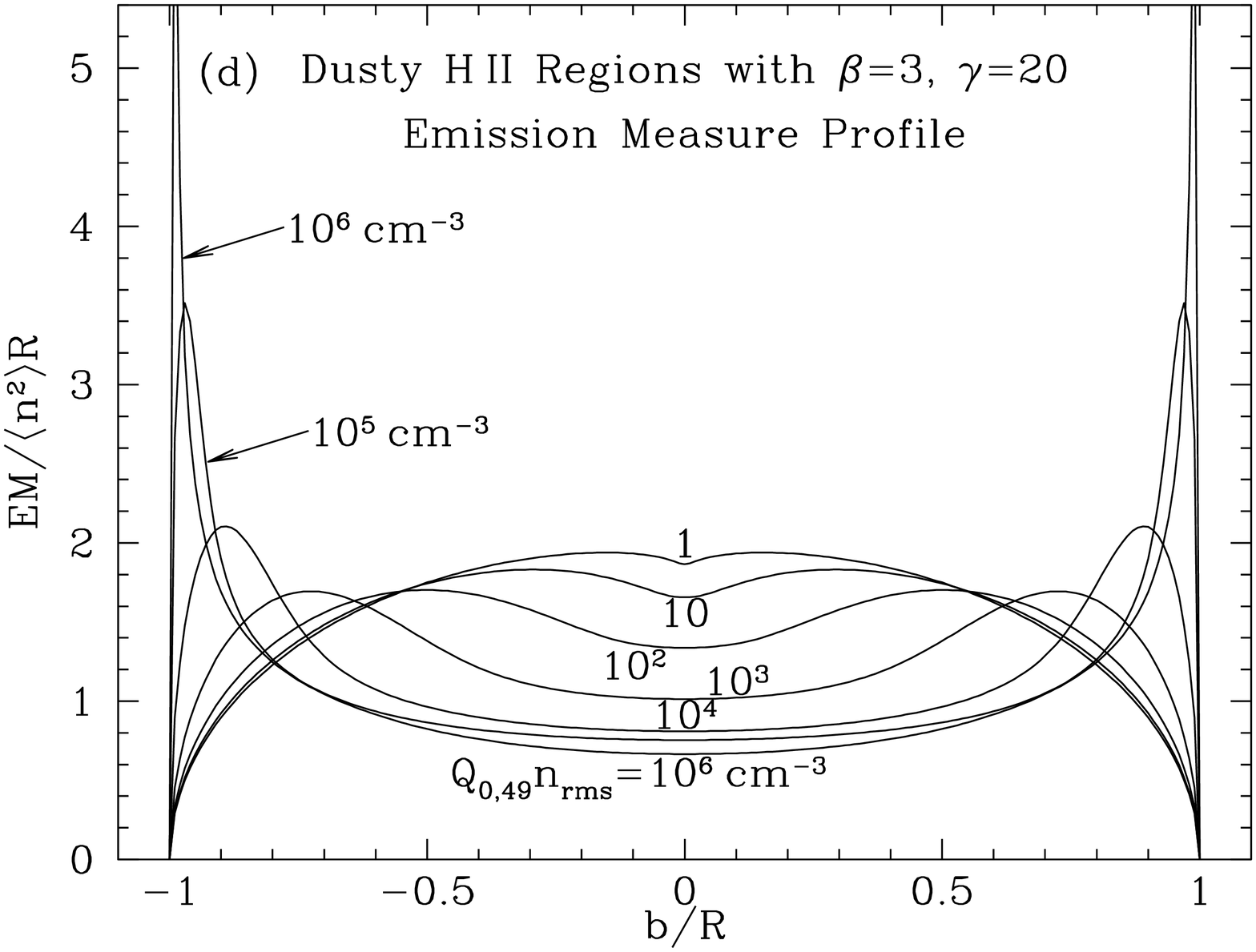}
\caption{\label{fig:Iprofs gamma=0, 5, 10, 20}
         \capsize
         Normalized emission measure (EM) profiles for a cut
         across the center of \ion{H}{2} regions with (a) $\gamma=0$ (no dust), 
         (b) $\gamma=5$, (c) $\gamma=10$, and (d) $\gamma=20$.
         Profiles are shown for selected values of $Q_0\nrms$.
         Numerical values of $Q_{0,49}\nrms$ assume $T_4=0.94$
         and $\langle h\nu\rangle_i=18\eV$.
         }
\end{center}
\end{figure}

The state of ionization of the gas is determined by the hardness of the
radiation field, and the value of the dimensionless ``ionization parameter''
\beq
U \equiv \frac{n(h\nu > \IH)}{\nH} ~~~,
\eeq
where $n(h\nu>\IH)$ is the density of photons with $h\nu>\IH$.
Within an \ion{H}{2} region, the value of $U$ varies radially.
As a representative value,
we evaluate $U_{1/2}$, the value at the 
``half-ionization'' radius $R_{1/2}$, the radius within
which 50\% of the H ionizations and recombinations take place.\footnote{
Some authors \citep[e.g.,][]{Dopita+Fischera+Crowley+etal_2006}
use the volume-averaged ionization parameter $\langle U\rangle_V$.  
For a uniform density dustless H\,II region,
$\langle U\rangle_V=(81/256\pi)^{1/3}(\alpha_B^{2/3}/c)(Q_0\nrms)^{1/3}
= 2.83\, U_{1/2}$.}
In a
uniform density dustless \ion{H}{2} region, $R_{1/2} = 2^{-1/3}R_{S0}$
is the same as the half-mass radius, and
\beq \label{eq:Uhalf, no dust}
U_{1/2}^{\rm (no\,dust)} 
= 
\frac{\alpha_B^{2/3}}{(72\pi)^{1/3}c} \left(Q_0 \nrms\right)^{1/3} ~~~.
\eeq
For our present models,
\beq
U_{1/2} = \frac{Q_0}{4\pi \lambda_0^2 n_0 c} 
\frac{\phi(y_{1/2})u(y_{1/2})}{y_{1/2}^2}~~~,
\eeq 
where $y_{1/2}=R_{1/2}/\lambda_0$ is the value of $y$ within which
50\% of the H ionizations and recombinations take place.
Figure \ref{fig:U} shows
$U_{1/2}$ as a function of $Q_0\nrms$ for
static dusty \ion{H}{2} regions with radiation pressure,
for selected values of $\beta$ and $\gamma$.
For small values of $Q_0 \nrms$, dust and radiation pressure are negligible
and $U_{1/2}$ coincides with $U_{1/2}^{\rm (no\,dust)}$ 
(eq.\ \ref{eq:Uhalf, no dust}).
However, for large values of $Q_0\nrms$, $U_{1/2}$ 
falls below $U_{1/2}^{\rm (no\,dust)}$.
For $\gamma\approx 10$ -- corresponding to the
dust abundances that we consider to be likely for Galactic \ion{H}{2}
regions -- we see that $U_{1/2} \approx 0.07 \pm 0.02$ for
$Q_{0,49}\nrms \gtsim 10^4 \cm^{-3}$.

The emission measure $EM(b)=\int n_e^2 ds$ 
is shown 
as a function of impact parameter $b$
in Figure \ref{fig:Iprofs gamma=0, 5, 10, 20}.
For small values of $Q_0\nrms$, the intensity profile is close to the
semicircular profile of a uniform density sphere.
As $Q_0\nrms$ is increased, the profile becomes flattened, but,
if no dust is present ($\gamma=0$, 
Fig.\ \ref{fig:nprofs gamma=0, 5, 10, 20}a), the ionized gas only begins
to develop an appreciable 
central minimum for $Q_{0,49}\nrms\gtsim 10^{4.5}\cm^{-3}$.

When dust is present, however, the profiles are strongly affected.
For standard parameters $\beta=3,\gamma=10$, the emission measure shows
a pronounced central minimum for $Q_{0,49}\nrms \gtsim 10^3\cm^{-3}$,
with a peak-to-minimum ratio $>2$ for $Q_{0,49}\nrms \gtsim 10^4\cm^{-3}$.
As $Q_{0}\nrms$ is
increased,
the ionized gas becomes concentrated in a thin, dense
shell,
the peak intensity near the edge rises, and the
central emission measure changes from $EM(0)=2\nrms^2 R$ for small
$Q_0\nrms$ to
$EM(0)\rightarrow (2/3)\nrms^2 R$ as $Q_0\nrms\rightarrow\infty$.

\begin{figure}[htb]
\begin{center}
\includegraphics[width=14.0cm,angle=0,
                 clip=true,trim=0.0cm 0.0cm 0.0cm 0.0cm]%
   {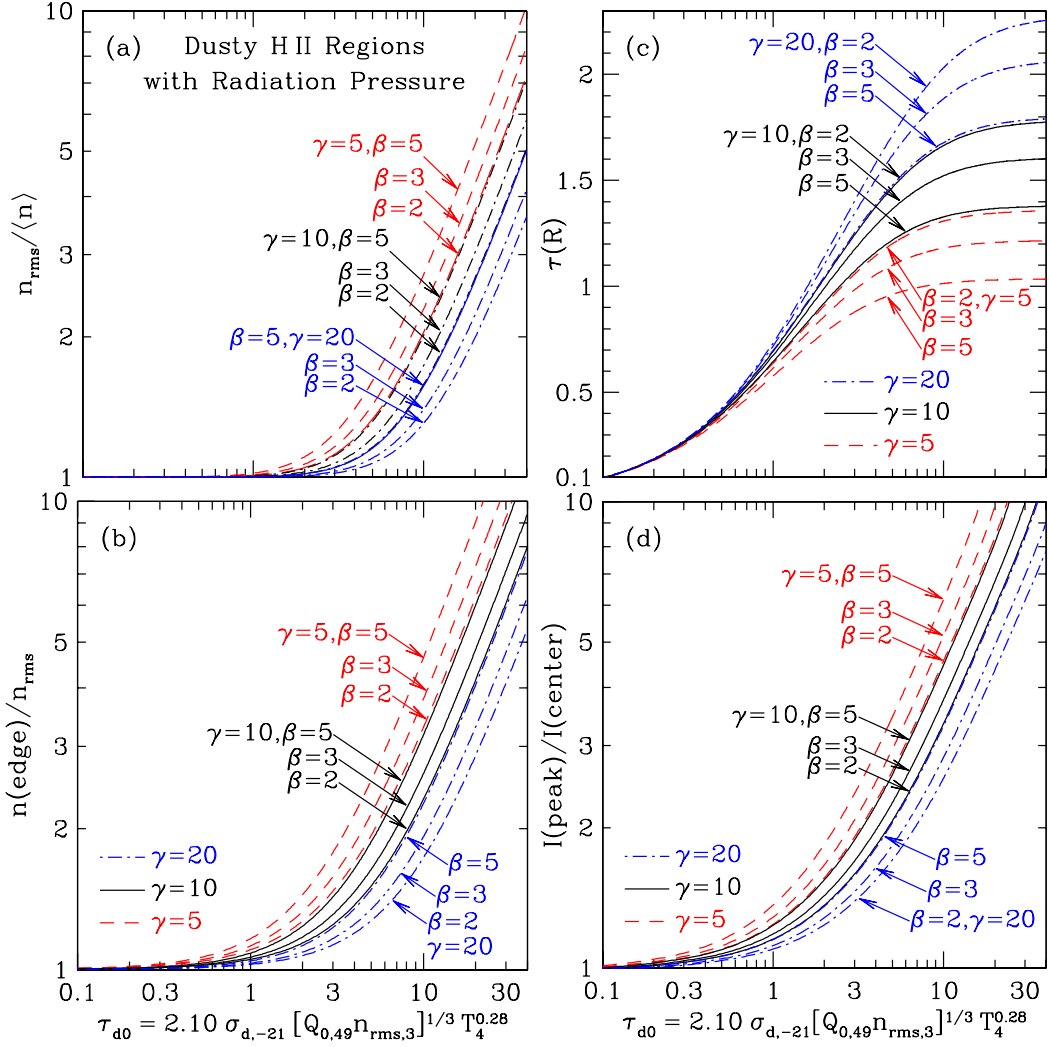}
\caption{\label{fig:model summ}
         \capsize
         For dusty \ion{H}{2} regions with $\gamma=5, 10, 20$ 
         and $\beta=2, 3, 5$, as a function of $\tau_{d0}$:
         (a) ratio $\nrms/\langle n\rangle$
         of rms density to mean density;
         (b) ratio $n(R)/\nrms$ of the edge density to the rms
         density;
         (c) center-to-edge dust optical depth $\tau(R)$;
         (d) ratio of peak emission measure/central emission measure.
         Results are for $\langle h\nu\rangle_i=18\eV$.
         }
\end{center}
\end{figure}
\begin{figure}[htb]
\begin{center}
\includegraphics[width=11cm,angle=0]%
   {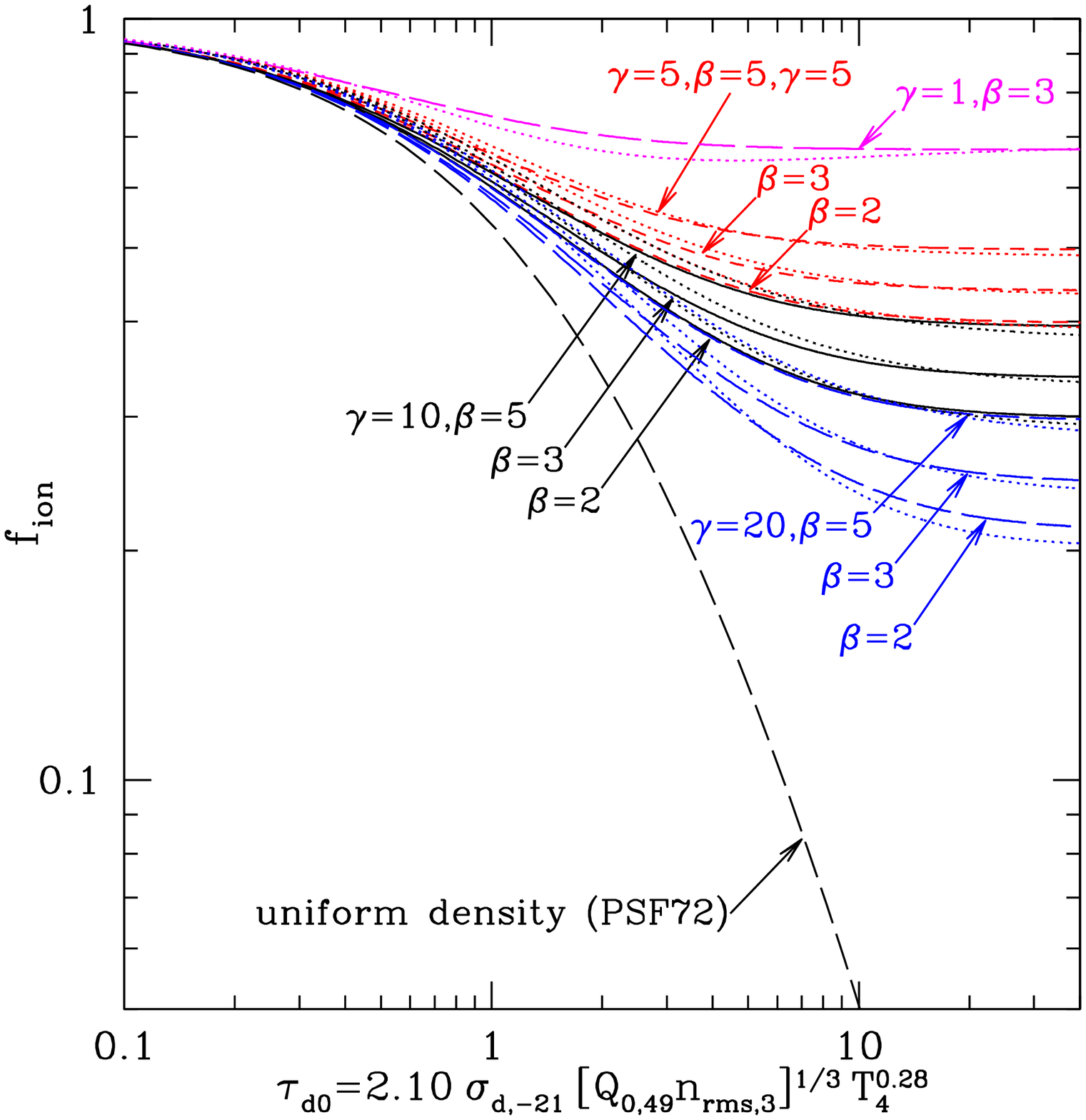}
\caption{\label{fig:log fion vs taud0}
         \capsize
         Fraction $\fion$ of the $h\nu>13.6\eV$ photons that 
         photoionize H in dusty \ion{H}{2} regions with
         radiation pressure, as a function of $\tau_{d0}$,
         for $\beta=2,3,5$ and $\gamma=5,10,20$.
         Calculated assuming $\langle h\nu\rangle_i=18\eV$.
         The dotted lines show the fitting formula
         (\ref{eq:fion fit}) for each 
         case.
         Also shown is $\fion$ calculated for assumed uniform density
         \citep{Petrosian+Silk+Field_1972}.
         }
\end{center}
\end{figure}

When $\tau_{d0}\gg1$, the present models have
the ionized gas concentrated in a thin, dense, shell.  
Figure \ref{fig:model summ}a shows the ratio of the rms density
$\nrms$
to the mean density $\langle n\rangle$ as a function of $\tau_{d0}$.
The highest ionized density occurs at the outer edge of the ionized
zone, and Figure \ref{fig:model summ}b shows the ratio
$n_{\rm edge}/\nrms$ as a function of $\tau_{d0}$.

In the low density limit $Q_0\nrms \rightarrow 0$, 
the dust optical depth from center to edge $\tau(R)\approx \tau_{d0}$.
The actual dust optical depth from center to edge is shown in
Figure \ref{fig:model summ}c.
When the \ion{H}{2} region develops a dense shell, which occurs for
$\tau_{d0}\gtsim 3$, the actual dust optical depth $\tau(R)$ is significantly
smaller than the value $\tau_{d0}$.  Figure \ref{fig:model summ}c shows
that for $\tau_{d0}=40$, for example, the actual dust optical depth
$\tau(R)$ is only in the range 1--2.3, depending on the values of $\beta$ and
$\gamma$.

The shell-like structure is also apparent in the ratio of the peak intensity
to the central intensity.  
As seen in Figures \ref{fig:Iprofs gamma=0, 5, 10, 20}(b-d), dust causes
the peak intensity to be off-center.
For fixed $\beta$ and $\gamma$, the ratio of peak intensity 
to central intensity $I({\rm peak})/I({\rm center})$
increases monotonically with increasing $\tau_{d0}$, as shown
in Fig.\ \ref{fig:model summ}(d).

Because the shell
is dense, radiative recombination is rapid, the neutral hydrogen
fraction is enhanced, and H atoms can compete with dust to absorb
$h\nu>13.6\eV$ photons.
Figure \ref{fig:log fion vs taud0} shows $\fion$, 
the fraction of the $h\nu>13.6\eV$
photons emitted by the star that photoionize H (i.e., are not absorbed
by dust), as a function of the parameter $\tau_{d0}$.  
Results are shown for $\beta=2, 3, 5$ and
$\gamma=5, 10, 20$.
For $2\leq\beta\leq5$, $5\leq\gamma\leq20$, and $0\leq\tau_{d0}\leq40$,
the numerical results in Figure \ref{fig:log fion vs taud0}can be
approximated by the fitting formula
\beqa \label{eq:fion fit}
\fion(\beta,\gamma,\tau_{d0}) 
&\approx& 
\frac{1}{1+(2/3+AB)\tau_{d0}} + \frac{AB\tau_{d0}}{1+B\tau_{d0}}
\\ \label{eq:define A}
A &=& \frac{1}{1+0.75\gamma^{0.65}\beta^{-0.44}}
\\ \label{eq:define B}
B &=& \frac{0.5}{1+0.1(\gamma/\beta)^{1.5}}
\eeqa
where $\beta$, $\gamma$, and $\tau_{d0}$ are given by
(\ref{eq:define beta}), (\ref{eq:define gamma}), and
(\ref{eq:taud0}).
The form of eq.\ (\ref{eq:fion fit}-\ref{eq:define B}) 
has no physical significance, but
eq.\ (\ref{eq:fion fit}) can be used to estimate the total 
H ionization rate $\fion Q_0$ in dusty \ion{H}{2} regions.

Even for large
values of $\tau_{d0}$, Fig.\ \ref{fig:log fion vs taud0} shows that
$\sim$1/3 of the $h\nu>13.6\eV$ photons are
absorbed by hydrogen.  
This contrasts with the uniform-density models
of \citet{Petrosian+Silk+Field_1972}, where the fraction of the
$h\nu>13.6\eV$ photons that are absorbed by the gas goes to zero as
$\tau_{d0}$ becomes large.

\section{\label{sec:dust drift}
          Dust Drift}
\subsection{Gas Drag vs.\ Radiation Pressure}
Eq.\ (\ref{eq:dynamical equilibrium})
assumes the dust to be tightly coupled to the gas,
so that the radiation pressure force on the dust can be considered to
act directly on the gas.
In reality,
radiation pressure will drive the dust grains through the plasma.
If the grains approach their terminal velocities (i.e., acceleration
can be neglected) then, as before, 
it can be assumed that the radiation pressure
force is effectively applied to the gas.  However, the motion of the
dust grains will lead to changes in the dust/gas ratio, due to
movement of the grains from one zone to another, as well as because of
grain destruction.  Here we estimate the drift velocities of grains.

\begin{figure}[htb]
\begin{center}
\includegraphics[width=8cm,angle=0]%
   {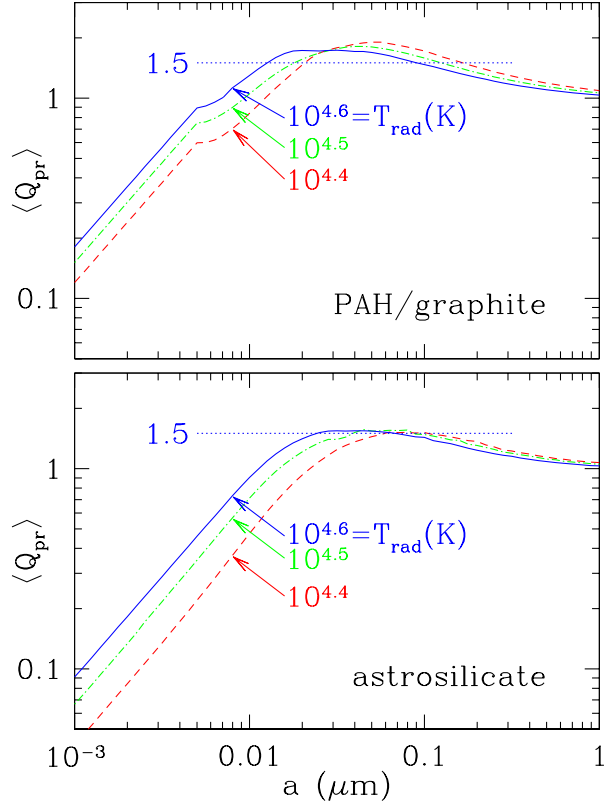}
\caption{\label{fig:Q_radpr vs. a}
   \capsize
   Spectrum-averaged radiation pressure efficiency factor
   $\langle Q_\radpr\rangle$ as a function of radius,
   for $T=25000\K$, 32000~K, and 40000~K blackbody spectra.
   For temperatures characteristic of O stars,
   $\langle Q_\radpr\rangle\approx 1.5$ to within 20\%
   for $0.02\micron \ltsim a \ltsim 0.25\micron$.
   }
\end{center}
\end{figure}

Let $Q_\radpr\pi a^2$ be the radiation pressure cross section
for a grain of radius $a$.
Figure \ref{fig:Q_radpr vs. a} shows $Q_\radpr(a,\lambda)$ averaged over
blackbody radiation fields with $T=25000\K$, 32000\,K, and 40000\,K,
for carbonaceous grains and amorphous silicate grains.
For spectra characteristic of O stars, $\langle Q_\radpr\rangle \approx 1.5$
for $0.02\micron\ltsim a \ltsim 0.25\micron$.

If the magnetic field $B=0$, the terminal velocity $v_d$ for a grain
at distance $r$ 
is determined by balancing the forces due to radiation pressure
and gas drag:
\beqa
\frac{L(r)}{4\pi r^2 c}\pi a^2 \langle Q_\radpr\rangle
&=&
2\pi a^2 nkT\,G(s) ~~~~,~~~~ s \equiv \frac{v_d}{\sqrt{2kT/\mH}} ~~~,
\eeqa
where the drag function $G(s)$,
including both collisional drag and Coulomb drag,
can be approximated by \citep{Draine+Salpeter_1979a}
\beqa
G(s) &\approx& \frac{8s}{3\sqrt{\pi}}\left(1+\frac{9\pi}{64}s^2\right)^{1/2}
+ \left(\frac{eU}{kT}\right)^2 \ln\Lambda \frac{s}{(3\sqrt{\pi}/4+s^3)}
~~~,~~~
\\
\Lambda&=&\frac{3}{2ae}\frac{kT}{|eU|}\left(\frac{kT}{\pi n_e}\right)^{1/2}
= 6.6\times10^6 a_{-5}^{-1} \frac{kT}{|eU|} T_4^{1/2} n_3^{-1/2}
~~~,~~~
\eeqa
where $U$ is the grain potential, $a_{-5}\equiv a/10^{-5}\cm$,
and $n_3\equiv n_e/10^3\cm^{-3}$.
The drag force from the electrons is smaller than that from
the ions by at least $\sqrt{m_e/m_p}$, and can be neglected.
The charge on the grains will be determined by collisional charging
and photoelectric emission.  Collisional charging would result in
$eU/kT\approx-2.51$ \citep{Spitzer_1968}, or $U\approx-2.16 T_4\Volt$.
Photoelectric charging will dominate close to the star, but is expected to
result in potentials $U\ltsim10\Volt$.
Taking
$|eU/kT|\approx 2.5$ and $\ln\Lambda\approx14.8$
as representative,
\beq
G(s)\approx \left[1.50\left(1+\frac{9\pi}{64}s^2\right)^{1/2}
+\frac{69.5}{1+4s^3/3\sqrt{\pi}}\right]s ~~~.
\eeq
Note that $G(s)$ is not monotonic: as $s$ increases from 0,
$G(s)$ reaches a peak value $\sim42$ for $s\approx 0.89$, but then
begins to decline with increasing $s$ as the Coulomb drag 
contribution falls.
At sufficiently large $s$, the direct collisional term becomes large
enough that $G(s)$ rises above $\sim42$ and continues to rise thereafter.

\begin{figure}[htb]
\begin{center}
\includegraphics[width=6cm,angle=270]%
   {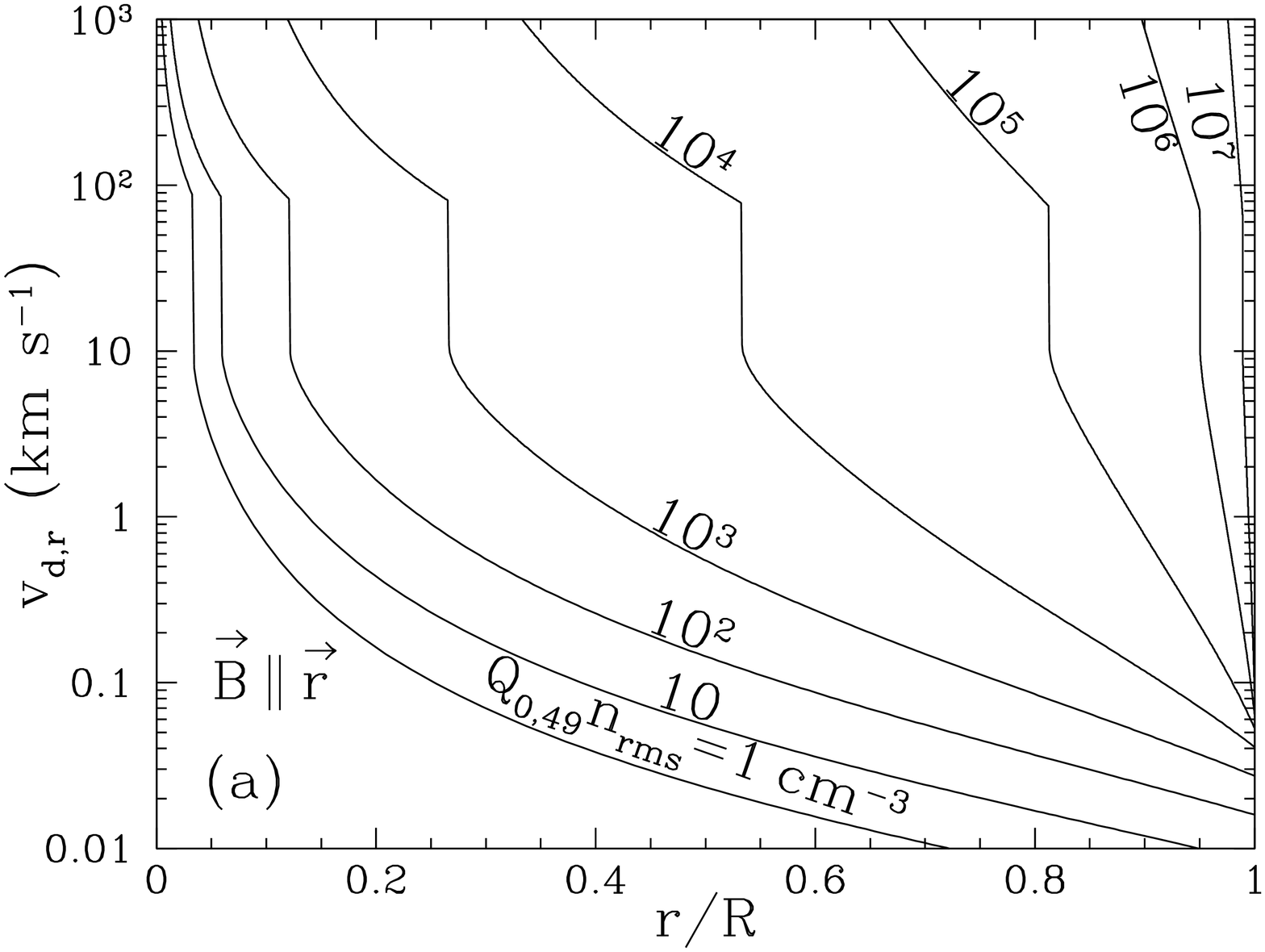}
\includegraphics[width=6cm,angle=270]%
   {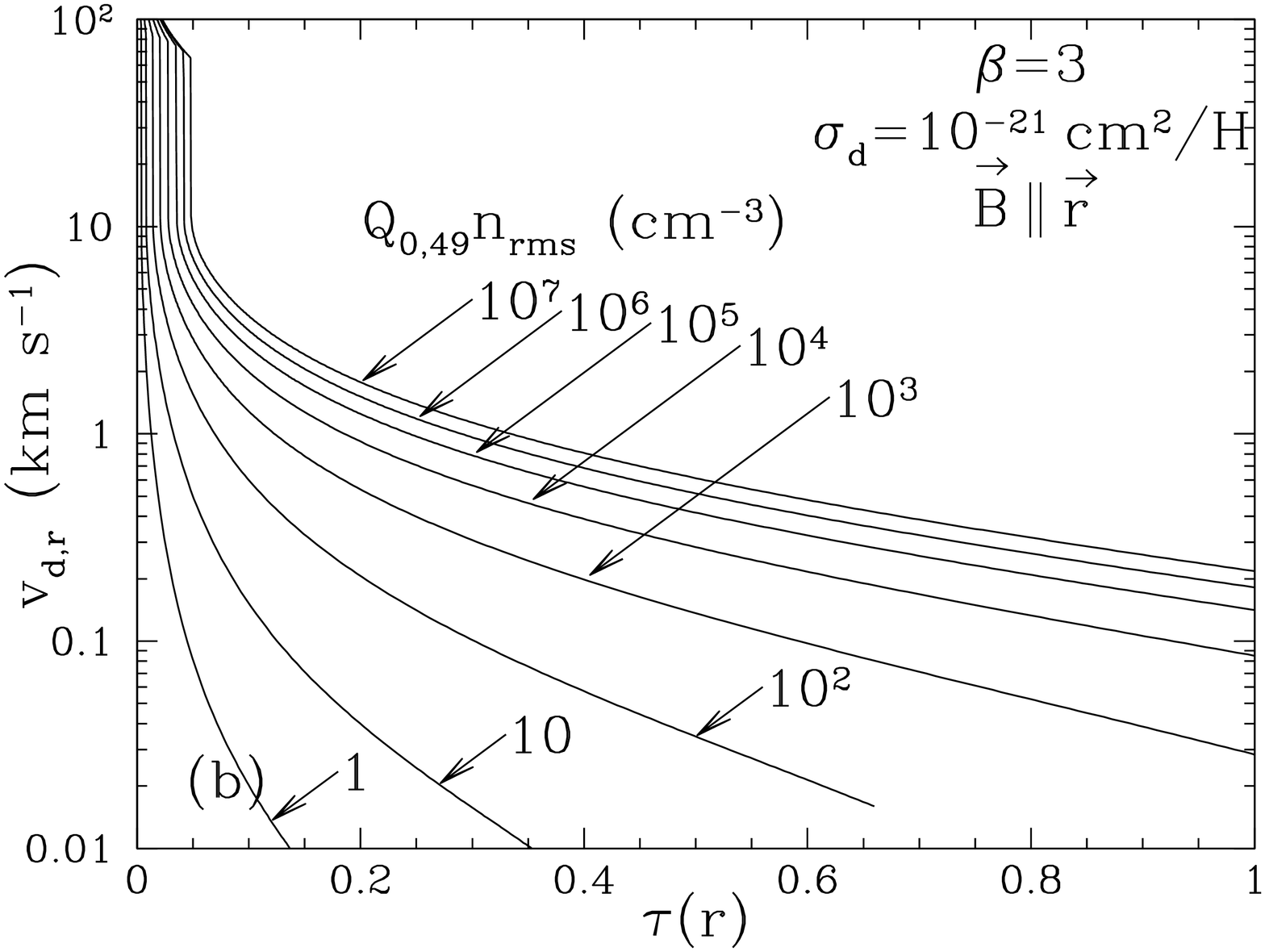}
\caption{\label{fig:vdrift gamma=10}
         \capsize
         Radial drift velocities $v_{d,r}$ for six
         different \ion{H}{2} regions, all with $\beta=3$ and
         $\gamma=10$, for $T_4=0.94$, $\langle h\nu\rangle_i=18\eV$,
         $Q_\radpr=1.5$, $|eU/kT|=2.5$, and $B=0$.
         (a) $v_{d,r}$ vs $r/R$.    
         All solutions have large drift velocities
         near the center, which will result in removal of the dust
         from the interior.
         Drift velocities increase with increasing $Q_0\nrms$.
         (b) $v_{d,r}$ as a function of dust column density
   $\tau(r)$.
   Even if $B=0$ or $\bB\parallel\br$, 
   drift velocities $v_d>75\kms$ occur only in a region
   with a small fraction of the dust.
   In most of the volume, drift will not result in grain destruction.
          }
\end{center}
\end{figure}

The drag time for a grain of density
$\rho=3\g\cm^{-3}$ in \ion{H}{2} gas is
\beq
\tau_{\rm drag} = \frac{Mv}{F_{\rm drag}}
= 295 \left(\frac{a_{-5}}{n_3 T_4^{1/2}}\right) ~ \frac{s}{G(s)}\yr ~~~.
\eeq
For $n_3\gtsim0.01$ this is sufficiently short that each grain
can be assumed to be moving at its terminal velocity $v_d$, with
isothermal Mach number
$s\equiv v_d/\sqrt{2kT/\mH}$ 
determined by the dimensionless equation
\beq \label{eq:eq for G(s)}
G(s) = 
\left[\phi(y)+\beta e^{-\tau(y)}\right]\frac{u(y)}{y^2}\langle Q_\radpr\rangle
~~~,~~~ y\equiv \frac{r}{\lambda_0}
~~~.~~~
\eeq
Eq. (\ref{eq:eq for G(s)}) is solved to find
$s(r)$.
For $20 < G < 42$, there are three values of $s$ for which the drag
force balances the radiation pressure force.  The intermediate solution
is unstable; we choose the smaller solution,\footnote{
   This solution is physically relevant if
   the drift speed began with $s\ltsim0.9$
   and increased with time.}
which means that $s$ undergoes
a discontinuous jump from $\sim 0.9$ to $6.2$ at $G\approx 42$.
The resulting terminal velocity $v(r)$ is shown in Figure
\ref{fig:vdrift gamma=10} for  7 
values of $Q_{0,49}\nrms$.
The velocities in the interior can be very large, but the velocities
where most of the dust is located [$\tau(r)/\tau(R)>0.5$] are much smaller.

In the outer part of the bubble, where most of the gas and dust
are located, the drift velocities are much more
modest.  This is visible in Fig.\ \ref{fig:vdrift gamma=10}a, where the
drift speeds become small as $r\rightarrow R$, but is more
clearly seen in Fig.\ \ref{fig:vdrift gamma=10}b, showing drift speeds
as a function of normalized optical depth.
The range
$0.5 < \tau(r)/\tau(R) < 1$ contains more than 50\% of the dust, and
throughout this zone the drift speeds are $\ltsim 0.3\kms$ even for 
$Q_{0,49}\nrms$ as large as $10^7\cm^{-3}$.
With drift speeds $v_d\ltsim 0.3\kms$, grains will not be destroyed, except
perhaps by shattering in occasional collisions between grains with
different drift speeds.  However, for large values of
$\nrms$, these grains 
are located close to the boundary, the drift
times may be short, and the
grains may be driven out of the \ion{H}{2} and into
the surrounding shell of dense neutral gas.
This will be discussed further below.

\subsection{Magnetic Fields}

Let
$\epsilon_B \equiv B^2/16\pi nkT$ be the ratio of magnetic pressure to
gas pressure.
The importance of magnetic fields for the grain dynamics
is determined by the dimensionless ratio 
$\omega \tau_{\rm drag}$, where $\omega\equiv QB/Mc$ is the gyrofrequency
for a grain with charge $Q$ and mass $M$ in a magnetic field $B$:
\beq \label{eq:omega tau}
\left(\omega\tau_{\rm drag}\right)^2 =
17.3 \frac{T_4^2}{n_3 a_{-5}^2} 
\left(\frac{\epsilon_B}{0.1}\right)
\left(\frac{eU/kT}{2.5}\right)^2
\left(\frac{71}{G(s)/s}\right)^2
~~~,~~~\epsilon_B\equiv\left(\frac{B^2/8\pi}{2nkT}\right) ~~~.~~~
\eeq
If $|eU/kT|\approx2.5$ and $\ln\Lambda\approx15$, then $(G(s)/s)\approx71$ 
for
$s\ltsim 0.5$.

Let the local magnetic field be 
$\bB=B(\hat{\br}\cos\theta+\hat{\by}\sin\theta)$.
The steady-state drift velocity is
\beqa \label{eq:vdrift}
v_d &=& \left(\frac{F_{\rm rad}\taudrag}{M}\right)
\sqrt{\frac{1+(\omega\taudrag)^2\cos^2\theta}{1+(\omega\taudrag)^2}}
~~~,~~~
\\
v_{d,r} &=& \left(\frac{F_{\rm rad}\taudrag}{M}\right)
\frac{1+(\omega\taudrag)^2\cos^2\theta}{1+(\omega\taudrag)^2}
~~~,~~~
\\
v_{d,y} &=& \left(\frac{F_{\rm rad}\taudrag}{M}\right)
\frac{(\omega\taudrag)^2\sin\theta\cos\theta}{1+(\omega\taudrag)^2}
~~~,~~~
\\
v_{d,z} &=& -\left(\frac{F_{\rm rad}\taudrag}{M}\right)
\frac{\omega\taudrag\sin\theta}{1+(\omega\taudrag)^2}
~~~,~~~
\eeqa
where $v_{d,r}$, $v_{d,y}$, $v_{d,z}$ are the radial and transverse
components.
If $\sin\theta\rightarrow 0$, the magnetic field does not affect
the radiation-pressure-driven drift velocity, but in the limit
$\sin\theta\rightarrow1$ magnetic effects can strongly suppress
the radial drift if
$\omega\taudrag\gg 1$ and $\cos\theta\ll 1$.

The magnetic field strength is uncertain, but it is unlikely that the
magnetic energy density will be comparable to the gas pressure; hence
$\epsilon_B\ltsim 0.1$.
From eq.\ (\ref{eq:omega tau}) it is then apparent that if the magnetic field
is strong ($\epsilon_B\approx 0.1$), 
magnetic effects on the grain dynamics can be important
in low density \ion{H}{2} regions, but will not
be important for very high densities: 
$(\omega\tau_{\rm drag})^2 \ltsim 1$ for 
$n_3\gtsim 170 a_{-5}^{-2}\epsilon_B$.

\subsection{Drift Timescale}

\begin{figure}[tb]
\begin{center}   
\includegraphics[width=8cm,angle=0]%
   {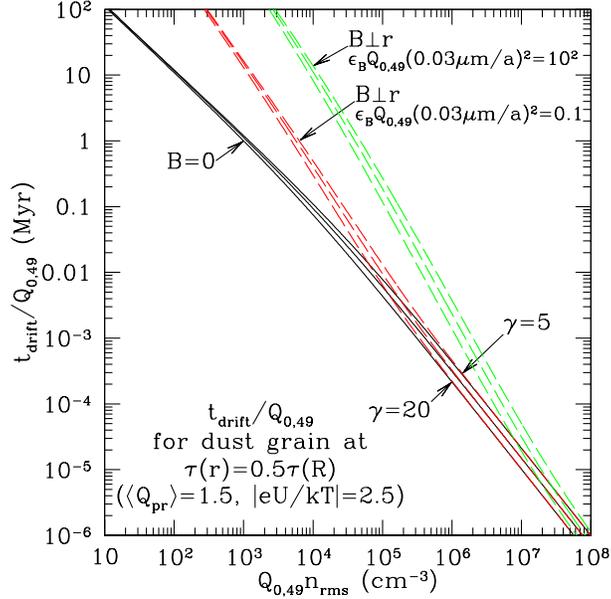}
\caption{\label{fig:tdrift}
   \capsize
   Drift timescale 
   $t_{\rm drift}/Q_{0,49}$ (see eq.\ \ref{eq:tdrift})
   for
   $\beta=3$ and $\gamma=5$, 10, and 20
   (assuming $T_4=0.94$, $\langle h\nu\rangle_i=18\eV$).
   The dust grains are assumed to have $\langle Q_\radpr\rangle=1.5$ 
   and $|eU/kT|=2.5$.  
   Solid lines are for $B=0$ (or $\bB\parallel\br$).
   Broken lines are for $a=0.03\micron$,
   and $\epsilon_B Q_{0,49}=0.1$ and $10^2$.
   }
\end{center}
\end{figure}

When radiation pressure effects are important, the gas and dust are
concentrated in a shell that becomes increasingly thin as
$Q_0 \nrms$ is increased.  The drift velocities where most of the
dust is located are not large (see
Fig.\ \ref{fig:vdrift gamma=10}b), but the grains are also not far from
the ionization front.
The timescale on which dust drift would be important can be estimated by
calculating the drift velocity at the radius $r_{0.5}$ defined by
$\tau(r_{0.5})=0.5\tau(R)$.  More than 50\% of the dust has $r_{0.5}<r<R$.
Figure \ref{fig:tdrift} shows the characteristic drift time
\beq \label{eq:tdrift}
t_{\rm drift} \equiv \frac{R-r_{0.5}}{v_{d,r}(r_{0.5})}
~~~.~~~
\eeq
If no magnetic field is present, the drift velocity depends only on
$T$ and the dimensionless quantities 
$\{\phi,\tau,u,y,\langle Q_\radpr\rangle\}$
(see eq.\ \ref{eq:eq for G(s)}).
It is easy to see that for fixed $T$ and $Q_0\nrms$, the radius
$R\propto Q_0$, thus
$t_{\rm drift}\propto Q_0$.
Figure \ref{fig:tdrift} shows $t_{\rm drift}/Q_{0,49}$.
For $Q_{0,49}=1$, \ion{H}{2} regions with $\nrms>10^3\cm^{-3}$ have
$t_{\rm drift}<10^6\yr$ if magnetic effects are negligible.
If a magnetic field is present with $\bB\perp\br$ and 
$\epsilon_B=0.1$, then the grain drift is slowed, but
drift times of $<1\Myr$ are found
for $\nrms > 10^4\cm^{-3}$.
Therefore, compact and ultracompact \ion{H}{2} regions around
single O stars are able to lower the dust/gas ratio by means of
radial drift of the dust on time scales of $\ltsim\Myr$.
However, if the O star is moving relative to the gas cloud with a
velocity of more than a few $\kms$, then individual fluid elements
pass through the ionized zone on timescales that may be shorter
than the drift timescale, precluding substantial changes in the
dust-to-gas ratio.

Grain removal by drift can also occur for giant \ion{H}{2} regions.
As an example, consider a giant \ion{H}{2} region ionized by
a compact cluster of $\sim$$10^3$ O stars 
emitting ionizing photons at a rate $Q_0=10^{52}\s^{-1}$.
For $\nrms=10^3\cm^{-3}$, we have $Q_{0,49}\nrms=10^6\cm^{-3}$,
and we see that if $B=0$, the drift timescale is only
$t_{\rm drift}\approx 2\times10^5\yr$.
If a magnetic field is present with $\bB\perp\br$ and
$\epsilon_B=0.1$, then from Figure \ref{fig:tdrift} the
drift timescale $t_{\rm drift}$ is increased, but only to $\sim10^6\yr$.
It therefore appears possible for radiation-pressure driven drift
to remove dust from giant \ion{H}{2} regions provided they are
sufficiently dense.

Aside from magnetic effects, the drift speeds at a given location
depend only on
$\langle Q_\radpr\rangle$ and $T_4$ (see eq.\ \ref{eq:eq for G(s)}).  
Figure \ref{fig:Q_radpr vs. a} shows
that $\langle Q_\radpr\rangle$ is constant to within a factor $\sim$1.5
for $a\gtsim 0.010\micron$.
Hence radiation-pressure-drive drift would act to drive
grains with $a\gtsim 0.01\micron$ outwards.
Smaller grains will drift as well, but more slowly.
Because of this, the gas-to-dust ratio in the centers of \ion{H}{2} regions
should in general be lower than the gas-to-dust ratio in the gas prior
to ionization.
The dust-to-gas ratio will first be reduced in the center, where
the drift speeds (see Fig.\ \ref{fig:vdrift gamma=10}) are large.
Dust drift will also alter the dust-to-gas ratio
in the outer ionized material, initially raising it by moving dust
outwards from the center.
In an initially uniform neutral cloud, the ionization front 
expands rapidly at early times
\citep[see, e.g., Fig. 37.3 in][]{Draine_2011a} but
in gas with $n_3\gtsim 1$, at late times the ionization front 
will slow to velocities small enough for dust grains to actually drift
outward across the ionization front, lowering the overall dust-to-gas
ratio within the \ion{H}{2} region.

\subsection{Grain Destruction}

\citet{Arthur+Kurtz+Franco+Albarran_2004} computed models of uniform density
\ion{H}{2} regions including the effects of dust destruction by
sublimation or evaporation, 
finding that the dust/gas ratio can be substantially reduced
near the star.
If the maximum temperature at which a grain can survive is $T_{\rm sub}$,
and the Planck-averaged absorption efficiencies are
$Q_{\rm uv}$ and $Q_{\rm ir}$ for $T=T_\star$ and $T=T_{\rm max}$, then grains
will be destroyed within a distance $r_{\rm sub}$ with
\beq
\frac{r_{\rm sub}}{R_{s0}} = 2.82\times10^{-3}L_{39}^{1/6}n_{\rm rms,3}^{2/3}
\left(\frac{10^3\K}{T_{\rm sub}}\right)^2
\left(\frac{L_{39}}{Q_{0,49}}\right)^{1/3}
\left(\frac{Q_{\rm uv}/Q_{\rm ir}}{10^2}\right)^{1/2}
\eeq
For parameters of interest (e.g., $L_{39}/Q_{0,49}\approx1$, 
$L_{39}\ltsim10^2$)
we find
$r_{\rm sub}/R_{s0}\ll 1$ for 
$\nrms\ltsim 10^5\cm^{-3}$,
and sublimation 
would therefore destroy only a small fraction of the dust.

As we have seen, we expect radiation pressure to 
drive grains through the gas, with velocity given by
eq.\ (\ref{eq:vdrift}). 
Drift velocities $v_d\gtsim75\kms$ will lead to sputtering by impacting
He ions, with sputtering yield $Y(\He)\approx 0.2$ for
$80\ltsim v \ltsim 500\kms$
\citep{Draine_1995b}.
For hypersonic motion, the grain of initial radius $a$ will be
destroyed after traversing a column density
\beq
\nH\Delta r = \frac{\nH}{n_{\rm He}}\frac{4\rho a}{Y(\He)\mu} 
\approx 
2\times10^{20}a_{-5}\left(\frac{Y(\He)}{0.2}\right)\cm^{-2}
\eeq
for a grain density $\rho/\mu=1\times10^{23}\cm^{-3}$,
appropriate for either silicates (e.g., FeMgSiO$_4$,
$\rho/\mu\approx3.8\g\cm^{-3}/25\mH=9\times10^{22}\cm^{-3}$)
or carbonaceous material ($2\g\cm^{-3}/12\mH=1.0\times10^{23}\cm^{-3}$).
Therefore the dust grain must traverse material with (initial) dust optical
depth
\beq \label{eq:Delta taud}
\Delta \tau_d = \sigma_d \nH\Delta r = 
0.2 \sigma_{d,-21} a_{-5} \left(\frac{Y(\He)}{0.2}\right)
\eeq
if it is to be substantially eroded by sputtering.
However, Fig.\ \ref{fig:vdrift gamma=10}b shows that even in the
absence of magnetic effects, $v_d\gtsim 75\kms$
occurs only in a central region with $\tau_d<0.05$.
Therefore sputtering arising from radiation-pressure-driven drift
will not appreciably affect the dust content.

\section{\label{sec:discussion}
         Discussion}

\subsection{Absorption of Ionizing Photons by Dust}


\begin{figure}[htb]
\begin{center}
\includegraphics[width=8cm,angle=0]%
   {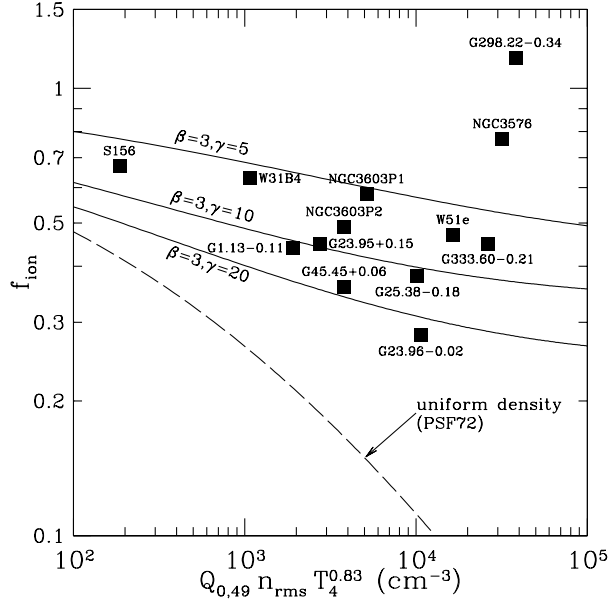}
\caption{\label{fig:inoue}
   \capsize
   Photoionizing fraction $\fion$ for 12 Galactic \ion{H}{2} regions,
   as estimated by \citet{Inoue_2002} from infrared and radio
   observations, vs.\ $Q_{0,49}n_e T_4^{0.83}$ (see text).
   $\fion$ cannot exceed 1, therefore the high value found for
   G298.22-0.34 give some indication of the uncertainties in estimation
   of $\fion$.
   Solid lines: $\fion$ for \ion{H}{2} regions with radiation 
   pressure for dust characterized by $\gamma=5$, 10, and 20.  
   Broken line: $\fion$ for uniform \ion{H}{2} regions with
   $\sigma_d=10^{-21}\cm^2\,{\rm H}^{-1}$.}
\end{center}
\end{figure}

For a sample of 13 Galactic \ion{H}{2} regions, \citet{Inoue_2002}
used infrared and radio continuum observations to obtain the
values of $\fion$ shown in 
Figure \ref{fig:inoue}.
The estimated values of $\fion$ are much larger than would be expected
for uniform \ion{H}{2} regions with dust-to-gas ratios comparable to the
values found in neutral clouds.
\citet{Inoue_2002} concluded that the central regions of
these \ion{H}{2} regions must be dust-free, noting that this was
likely to be due to the combined effects of stellar winds and
radiation pressure on dust.
As seen in Fig.\ \ref{fig:inoue}, the values of $\fion$ found by
Inoue are entirely consistent with what is expected for static
\ion{H}{2} regions
with radiation pressure for $5 \ltsim \gamma \ltsim 20$
(corresponding to $0.5 \ltsim \sigma_{d,-21}\ltsim 2$), with no need
to appeal to stellar winds or grain destruction.

\subsection{The Density-Size Correlation for \ion{H}{2} Regions}

\begin{figure}[htb]
\begin{center}
\includegraphics[width=8cm,angle=0]%
   {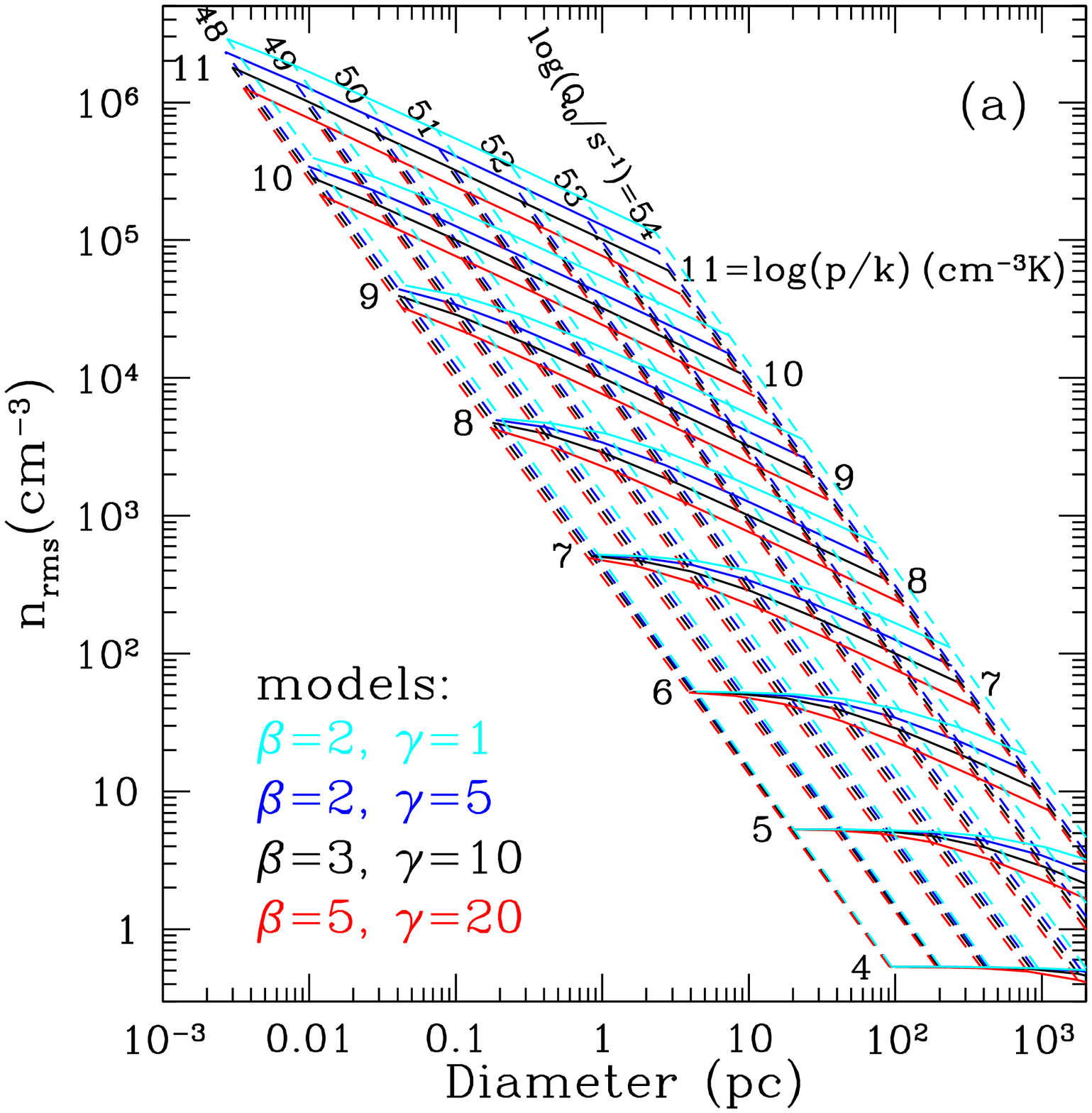}
\includegraphics[width=8cm,angle=0]%
   {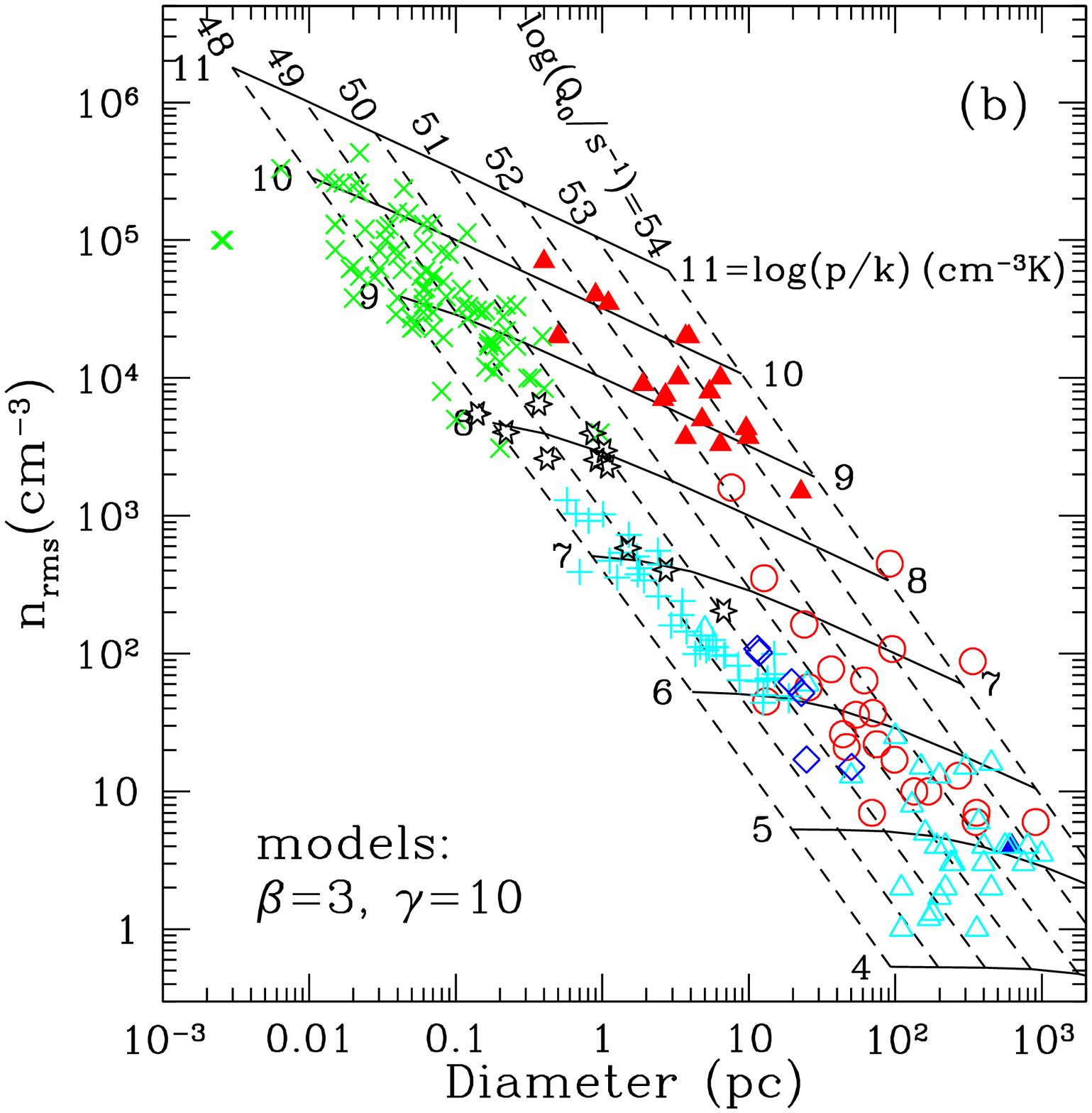}
\caption{\label{fig:ne vs D}
   \capsize
   Density $\nrms$ vs.\ diameter $D$.
   (a) Models with
   $(\beta,\gamma)=$ (2,1), (2,5), (3,10), and (5,20).
   Results shown were calculated for $T_4=0.94$, $\langle h\nu\rangle_i=18\eV$.
   Solid lines show models with $p_{\rm edge}$ fixed, and
   $Q_0$ varying from $10^{48}\s^{-1}$ to $10^{54}\s^{-1}$.
   Broken lines show models with $Q_0$ fixed, and $p_{\rm edge}/k$
   varying from $10^4\cm^{-3}\K$ to $10^{11}\cm^{-3}\K$.
   (b)
   Model grid for $\beta=3$, $\gamma=10$ together with
   observed values are shown for various samples
   of Galactic and extragalactic H\,II regions.
   Cyan open triangles: \citet{Kennicutt_1984}.
   Blue diamonds: \citet{Churchwell+Goss_1999}.
   Green crosses: \citet{Garay+Lizano_1999}.
   Cyan crosses: \citet{Kim+Koo_2001}.
   Black open stars: \citet{Martin-Hernandez+Vermeij+vanderHulst_2005}.
   Red solid triangles: radio sample from \citet{Hunt+Hirashita_2009}.
   Red open circles: HST sample from \citet{Hunt+Hirashita_2009}.
   }
\end{center}
\end{figure}

\ion{H}{2} regions come in many sizes, ranging from \ion{H}{2}
regions powered by a single O star, to giant \ion{H}{2} regions ionized
by a cluster of massive stars.  The physical size of the \ion{H}{2}
region is obviously determined both by the total ionizing output
$Q_0$ provided by the ionizing stars, and the r.m.s.\ density $\nrms$ of
the ionized gas, which is regulated by the pressure $p_{\rm edge}$ of the
confining medium.
With the balance between photoionization and recombination determining
the size of an \ion{H}{2} region, an anticorrelation between size $D$
and density $\nrms$ is expected, and was observed as soon as large samples of
\ion{H}{2} regions became available
\citep[e.g.,][]{Habing+Israel_1979,Kennicutt_1984}.
For dustless \ion{H}{2} regions, one expects
$\nrms\propto D^{-1.5}$ for fixed $Q_0$, but for various samples
relations close to
$\nrms\propto D^{-1}$ were reported
\citep[e.g.,][]{Garay+Rodriguez+Moran+Churchwell_1993,
       Garay+Lizano_1999,
       Kim+Koo_2001,
       Martin-Hernandez+Vermeij+vanderHulst_2005}.
For ultracompact \ion{H}{2} regions,
\citet{Kim+Koo_2001} attribute the $n_e\propto D^{-1}$ trend
to a ``champagne flow'' and the hierarchical structure of the dense
gas in the star-forming region, but
\citet{Arthur+Kurtz+Franco+Albarran_2004} and
\citet{Dopita+Fischera+Crowley+etal_2006} argue that
the $n_e\propto D^{-1}$ trend is
a result of both absorption by dust and radiation
pressure acting on dust in static \ion{H}{2} regions.

\citet{Hunt+Hirashita_2009} recently reexamined the size-density 
relationship.  They interpreted the size-density relation for different
observational samples in terms of models with different star formation rates
[and hence different time evolution of the ionizing output $Q_0(t)$],
and differences in the density of the neutral cloud into which the
\ion{H}{2} region expands.
Their models did not include the effects of radiation pressure on dust;
at any time the ionized gas in an \ion{H}{2} region
was taken to have uniform density, resulting in overestimation of
the dust absorption.

Figure \ref{fig:ne vs D}a shows a grid of $\nrms$ vs.\ $D$
for the present models, for
four combinations of $(\beta,\gamma)$.
While differences between the models with different $(\beta,\gamma)$
can be seen, especially for high $Q_0$ and high $p_{\rm edge}$,
the overall trends are only weakly dependent on $\beta$ and $\gamma$, at
least for $1\ltsim\gamma\ltsim 20$.

Figure \ref{fig:ne vs D}b shows the model grid for
$\beta=3$ and $\gamma=5$ together with observed values of $D$ and $\nrms$
from a number of different studies.
It appears that observed \ion{H}{2} regions -- ranging from
\ion{H}{2} regions ionized by one or at most a few O stars
($Q_0<10^{50}\s^{-1}$) to ``super star clusters'' powered by
up to $10^3-10^5$ O stars ($Q_0=10^{52}-10^{54}\s^{-1}$)
can be accomodated by the present static equilibrium models
with external pressures in the range 
$10^4
\ltsim p/k \ltsim 10^{10.3}\cm^{-3}\K$.
Note that for diameters $D\gtsim 10^2\pc$, the assumption of
static equilibrium is unlikely to be justified, because the sound-crossing
time $(D/2)/15\kms\gtsim 3 \Myr$ becomes longer than the lifetimes of 
high-mass stars.

The fact that some \ion{H}{2} region samples 
\citep[e.g.,][]{Garay+Rodriguez+Moran+Churchwell_1993,Kim+Koo_2001}
seem to obey a
$\nrms\propto D^{-1}$ relationship appears to be an artifact of the
sample selection.  We see in Fig.\ \ref{fig:ne vs D}b that the overall
sample of \ion{H}{2} regions does not have a single $\nrms$-vs.-$D$
relationship.  But the observations appear to be generally consistent
with the current models of dusty \ion{H}{2} regions.

\subsection{Cavities in \ion{H}{2} Regions: N49}

Even without dust present, radiation pressure from photoelectric absorption
by H and He
can alter the density profile in a static \ion{H}{2} region, lowering the
central density and enhancing the density near the edge of the ionized
region (see Fig.\ \ref{fig:nprofs gamma=0, 5, 10, 20}a).  As seen in
Figure \ref{fig:Iprofs gamma=0, 5, 10, 20}a, for large values
of $Q_0\nrms$ the surface brightness profile
can be noticeably flattened.
If dust is assumed to be present, with properties typical of the dust
in diffuse clouds, the equilibrium density profile changes dramatically,
with a central cavity surrounded by a high-pressure shell of ionized gas
pushed out by radiation pressure.
In real \ion{H}{2} regions, fast stellar winds will also act to inflate
a low-density cavity, or ``bubble'', near the star; the observed
density profile will be the combined result of the stellar wind bubble
and the effects of radiation pressure.

The GLIMPSE survey \citep{Churchwell+Babler+Meade+etal_2009} has 
discovered and catalogued numerous interstellar ``bubbles''.
An example is N49 \citep{Watson+Povich+Churchwell+etal_2008}, with
a ring of free-free continuum emission at 20~cm, surrounded by a ring of
$8\micron$ PAH emission.
An O6.5V star is located near the center of the N49 ring.
The image is nearly circularly symmetric, with only a modest asymmetry
that could be due to motion of the star relative to the gas.
The 20~cm image has a ring-peak-to-center
intensity ratio $I({\rm peak})/I({\rm center})\approx 2$.

Is the density profile in N49 consistent with what is expected
for radiation pressure acting on dust?
From the 2.89~Jy flux from N49 at $\lambda=20$~cm 
\citep{Helfand+Becker+White+etal_2006} 
and distance $5.7\pm0.6\kpc$
\citep{Churchwell+Povich+Allen+etal_2006}, the stellar source 
has $Q_{0,49}\approx (0.78\pm0.16)/\fion$.
If $\fion\approx 0.6$, then  $Q_{0,49}\approx (1.3\pm0.3)$.
The \ion{H}{2} region, with radius $(0.018\pm0.02)$~deg,
has $\nrms\approx197\pm63\cm^{-3}$.
Hence
$Q_{0,49}\nrms\approx 260\cm^{-3}$.
If $\sigma_{d,-21}=1$, then $\tau_{d0}\approx 1.3$.
From Fig.\ \ref{fig:log fion vs taud0}a we confirm that $\fion\approx 0.6$ for
$\tau_{d0}\approx 1.3$.

Figure \ref{fig:model summ}d shows that 
an \ion{H}{2} region with $\tau_{d0}=1.3$ is expected to have
a central minimum in the emission measure, but with
$I({\rm peak})/I({\rm center})\approx 1.3$ for $\beta=3,\gamma=10$,
whereas the observed
$I({\rm peak})/I({\rm center})\approx 2$.
The central cavity in N49 is therefore significantly 
larger than would be expected
based on radiation pressure alone.
While the effects of radiation pressure are not negligible in N49,
the observed 
cavity must be the result of the combined effects of radiation pressure
and a
dynamically-important stellar wind (which is of course not unexpected
for an O6.5V star).

\subsection{Lyman-$\alpha$}

The original ionizing photon deposits a radial momentum $h\nu_i/c$ at the
point where it is absorbed by either a neutral atom or a dust grain.
A fraction ($1-\fion$) of the ionizing photons 
are absorbed by dust; this energy is 
reradiated isotropically, with  
no additional force exerted on the emitting material.
Because the infrared optical depth within the \ion{H}{2} region is small,
the infrared emission escapes freely, with no dynamical effect within the
\ion{H}{2} region.

A fraction $\fion$ of the ionizing energy is absorbed by the gas.
Subsequent radiative 
recombination and radiative cooling converts this energy to photons, but
the isotropic emission process itself
involves no net momentum transfer to the gas.
We have seen above that the \ion{H}{2} can have a center-to-edge 
dust optical depth
$\tau(R)\approx 1.6$ for $\tau_{d0}\gtsim 5$,
or $Q_{0,49}\nrms\gtsim10^{2}\cm^{-3}$ (cf.\ Fig.\ \ref{fig:model summ}c 
with $\beta=3$, $\gamma=10$).
This optical depth applies to the $h\nu>13.6\eV$ ionizing radiation;
the center-to-edge
optical depth for the $h\nu<3.4\eV$ Balmer lines and collisionally-excited
cooling lines emitted by the ionized gas will be significantly smaller,
and much of this
radiation will escape dust absorption or scattering 
within the \ion{H}{2} region.
That which is absorbed or scattered will exert a force on the dust at that
point only to the extent that
the diffuse radiation field is anisotropic.
We conclude that momentum deposition from 
the Balmer lines and collisionally-excited cooling lines
within the ionized zone will be small compared to the momentum
deposited by stellar photons.

Lyman-$\alpha$ is a special case.
At low densities ($n \ll 10^3\cm^{-3}$), 
$\sim70\%$ of Case B recombinations result in emission of
a Ly-$\alpha$ photon, increasing to $>95\%$ for $n>10^5\cm^{-3}$
as a result of collisionally-induced $2s\rightarrow2p$ transitions
\citep{Brown+Mathews_1970}.
After being emitted isotropically, the 
photon may scatter many times before either escaping
from the \ion{H}{2} or being absorbed by dust.
Most of the scatterings take place near the point of emission,
while the photon frequency is still close to line-center.
On average, the net radial momentum transfer per emitted photon
will likely be dominated 
by the last scattering event before the photon escapes from the
\ion{H}{2} region, or by the dust absorption event if it does not.
At a given point in the nebula, the incident photons involved in these
final events will be only moderately anisotropic.
Since there is less than one 
Ly-$\alpha$ photon created per case B recombination,
the total radial momentum deposited by these final events will be
a small fraction of the radial momentum of the original ionizing photons.
\citet{Henney+Arthur_1998} estimate that dust limits the Ly-$\alpha$
radiation pressure to $\sim$$6\%$ of the gas pressure.
We conclude that Ly-$\alpha$ has only a minor effect on the density
profile within the ionized zone.

\subsection{\ion{H}{2} Region Expansion}
 
\ion{H}{2} regions arise when massive stars begin to emit ionizing radiation.
The development of the \ion{H}{2} region over time depends on the growth of the
ionizing output from the central star, and the expansion of the
initially-high pressure ionizing gas.
Many authors \citep[e.g.,][]{Kahn_1954,Spitzer_1978} 
have discussed the development
of an \ion{H}{2} region in gas that is initially neutral and uniform.
If the ionizing output from the star turns on suddenly, 
the ionization front is initially
``strong R-type'', propagating supersonically without affecting the
density of the gas, slowing until it
becomes ``R-critical'', at which point it
makes a transition to ``D-type'', 
with the ionization front now preceded by a shock wave producing
a dense (expanding) shell of neutral gas bounding the ionized region.

While the front is R-type, the gas density and pressure are essentially
uniform within the ionized zone.
When the front becomes D-type, a rarefaction wave propagates inward from
the ionization front, but the gas pressure (if radiation pressure effects
are not important) remains
relatively uniform within the ionized region, because the motions
in the ionized gas are subsonic.

When radiation pressure effects are included, the instantaneous density profile
interior to the ionization front is expected to be similar to the profile 
calculated for the static equilibria studied here.
Let $V_i$ be the velocity of the ionization front relative
to the star.
When the ionization front is weak D-type, the velocity of the ionization front
relative to the ionized gas just inside the ionization front is
$\sim0.5 V_i$ \citep{Spitzer_1978}.
Given the small dust drift velocities $v_{d,r}$ near the ionization front
(i.e., $\tau(r)\rightarrow \tau(R)$ in Fig.\ \ref{fig:vdrift gamma=10}),
dust is unable to drift outward across the ionization front
as long as the ionization front is propagating outward with a speed
(relative to the ionized gas) $V_i\gtsim 0.1\kms$

\section{\label{sec:summary}
         Summary}
\begin{enumerate}
\item
Dusty \ion{H}{2} regions in static equilibrium 
consist of a three-parameter family of similarity solutions, 
parametrized by parameters $\beta$, $\gamma$, and a 
third parameter, which can be
taken to be $Q_{0,49}\nrms$ or $\tau_{d0}$ (see eq.\ \ref{eq:taud0}).
The $\beta$ parameter (eq.\ \ref{eq:define beta})
characterizes the relative importance of
$h\nu<13.6\eV$ photons, and $\gamma$ (eq.\ \ref{eq:define gamma})
characterizes the dust opacity.
A fourth parameter -- e.g., the value of $\nrms$ or $Q_{0,49}$ --
determines the overall size and density of the \ion{H}{2} region.
\item Radiation pressure acting on both gas and dust can strongly affect 
the structure of \ion{H}{2} regions.
For dust characteristic of the diffuse ISM of the Milky Way, static 
\ion{H}{2} regions with $Q_{0,49}\nrms\ltsim 10^2\cm^{-3}$ will 
have nearly uniform
density, but when $Q_{0,49}\nrms\gg 10^2\cm^{-3}$, radiation pressure acts
to concentrate the gas in a spherical shell.
\item For given $\beta$ and $\gamma$, the importance of radiation pressure
is determined mainly by the parameter $\tau_{d0}$ (see eq.\ \ref{eq:taud0}).
When $\tau_{d0}\gtsim 1$, radiation pressure will produce a noticeable
central cavity.
\item If the dust-to-gas ratio is similar to the
value in the Milky Way, then compression of the ionized gas into a shell
limits the
characteristic ionization parameter: $U_{1/2}\ltsim 0.01$,
even for $Q_0 \nrms\gg 1$
(see Fig.\ \ref{fig:U}).
\item For $\tau_{d0}\gtsim 1$, compression of the gas and dust 
into an ionized shell leads to a substantial {\it increase}
\citep[compared to the estimate by][]{Petrosian+Silk+Field_1972} 
in the fraction $\fion$ of $h\nu>13.6\eV$ photons
that actually ionize H, relative to what would have been estimated for a uniform
density \ion{H}{2} region, as shown in Fig.\ \ref{fig:log fion vs taud0}.
Eq.\ (\ref{eq:fion fit}) allows $\fion$ to be estimated for given
$Q_0\nrms$, $\beta$, and $\gamma$.
Galactic \ion{H}{2} regions appear to have values of $\fion$ consistent
with the present results for \ion{H}{2} regions with radiation pressure
(see Fig.\ \ref{fig:inoue}).
\item Interstellar bubbles surrounding O stars are the result of
the combined effects of radiation pressure and stellar winds.
For the N49 bubble, as an example, the observed ring-like free-free
emission profile
is more strongly peaked than would
be expected from radiation pressure alone,
implying that a fast stellar wind must be present to help create the
low-density central cavity.
\item For static \ion{H}{2} regions, dust drift would be important on
time scales $\ltsim1\Myr$ for $Q_{0,49}\nrms\gtsim 10^3\cm^{-3}$.
Real \ion{H}{2} regions are not static, and the dust will not drift
out of the ionized gas because the ionization front will generally
be propagating (relative to the ionized gas just inside the ionization
front) faster than the dust drift speed $\ltsim 1\kms$
(see Fig.\ \ref{fig:vdrift gamma=10}).
\end{enumerate}
\acknowledgements
I am grateful to Bob Benjamin and Leslie Hunt for helpful discussions, to
R.H. Lupton for making available the SM graphics package, and
to the anonymous referee for suggestions that improved the paper.
This research made use of NASA's Astrophysics Data System Service,
and was supported in part by NASA through JPL contract 1329088,
and in part by NSF grant AST 1008570.

\bibliography{btdrefs}

\end{document}